\documentclass[twocolumn,showpacs,preprintnumbers,amsmath,amssymb,superscriptaddress]{revtex4}

\usepackage{amsmath,amsfonts,amssymb}
\usepackage[english]{babel} 
\usepackage[latin1]{inputenc} 
\usepackage[T1]{fontenc}
\usepackage{color}
\usepackage{float}
\usepackage{verbatim}

\usepackage{calc}
\usepackage{float}
\usepackage{epsfig} 
\usepackage{graphicx}
\usepackage{bm}
\def \equi#1{\mathrel{\mathop{\kern 0pt\sim}\limits_{#1}}} 
\scrollmode

\newcommand{\moy}[1]{\left\langle #1 \right\rangle}

\newcommand{\ex}[1]{\mathrm{e}^{#1}}
\newcommand{\epu}[0]{\mathbf{e_1}}

\newcommand{\emuu}[0]{\mathbf{e_{\boldsymbol{\mu}}}}

\newcommand{\enu}[0]{\mathbf{e_{\boldsymbol \nu}}}

\newcommand{\dd}[0]{\mathrm{d}}

\newcommand{\zz}[0]{\mathbf{0}}

\begin{document}


\title{Geometry-induced superdiffusion in driven crowded systems }

\author{Olivier B\'enichou}
\affiliation{Laboratoire de Physique Th\'eorique de la Mati\`ere Condens\'ee (UMR CNRS 7600), Universit\'e Pierre et Marie Curie, 4 Place Jussieu, 75255
Paris Cedex France}

\author{Anna Bodrova}
\affiliation{Department of Physics, Moscow State University, Moscow, Russia}

\author{Dipanjan Chakraborty}
\affiliation{Max-Planck-Institut f\"ur Intelligente Systeme, Heisenbergstr. 3, 70569 Stuttgart, Germany, and 
Institut f\"ur Theoretische Physik IV,
Universit\"at Stuttgart, Pfaffenwaldring 57, 70569 Stuttgart,
Germany}

\author{Pierre Illien}
\affiliation{Laboratoire de Physique Th\'eorique de la Mati\`ere Condens\'ee (UMR CNRS 7600), Universit\'e Pierre et Marie Curie, 4 Place Jussieu, 75255
Paris Cedex France}

\author{Adam Law}
\affiliation{Max-Planck-Institut f\"ur Intelligente Systeme, Heisenbergstr. 3, 70569 Stuttgart, Germany, and 
Institut f\"ur Theoretische Physik IV,
Universit\"at Stuttgart, Pfaffenwaldring 57, 70569 Stuttgart,
Germany}

\author{Carlos Mej\'{\i}a-Monasterio}
\affiliation{Laboratory of Physical Properties,
Technical University of Madrid, Av. Complutense s/n, 28040 Madrid, Spain, and
Department of Mathematics and Statistics,
    University of  Helsinki, P.O.  Box 68  FIN-00014 Helsinki, Finland}

\author{Gleb Oshanin}
\affiliation{Laboratoire de Physique Th\'eorique de la Mati\`ere Condens\'ee (UMR CNRS 7600), Universit\'e Pierre et Marie Curie, 4 Place Jussieu, 75255
Paris Cedex France}

\author{Rapha\"el Voituriez}
\affiliation{Laboratoire de Physique Th\'eorique de la Mati\`ere Condens\'ee (UMR CNRS 7600), Universit\'e Pierre et Marie Curie, 4 Place Jussieu, 75255
Paris Cedex France}
\affiliation{Laboratoire Jean Perrin, FRE 3231 CNRS /UPMC, 4 Place Jussieu, 75255
Paris Cedex}

\date{\today}

\begin{abstract}

Recent Molecular Dynamics  simulations  of  glass-forming liquids  revealed superdiffusive fluctuations associated with the position of a tracer particle (TP) driven by an external force. Such anomalous response, whose mechanism remains elusive,  has been observed up to now  only in systems close to their glass transition,  suggesting that this could be one of its hallmarks. Here, we show that the presence of superdiffusion is in actual fact much more general, provided that the system is crowded and geometrically confined. We present and solve analytically a minimal model consisting of a driven TP  in a dense, crowded medium in which the motion of particles is mediated by the diffusion of packing defects, called vacancies. For such non glass-forming systems, our analysis predicts a long-lived superdiffusion which ultimately crosses over to giant diffusive behavior. We find that this trait is  present in confined geometries, for example long capillaries and stripes, and  emerges as a universal response of crowded environments to an external force. These findings are confirmed by numerical simulations of systems as varied as lattice-gases, dense  liquids and granular fluids. 
 \end{abstract}

\pacs{ }

\maketitle


Active microrheology monitors the response of a medium whilst in the presence of a tracer particle (TP) manipulated by an external force.  It has become a powerful experimental tool for the analysis of different systems such as colloidal suspensions \cite{Squires:2005,Zia:2013,Habdas:2004,Meyer:2006,Wilson:2009}, glass forming liquids \cite{Jack:2008,Winter:2012,Harrer:2012,Winter:2013,Schroer:2013}, fluid interfaces  \cite{Choi:2011} or live cells \cite{Valberg01071985,HEnon:1999}.  It constitutes a striking realization of a key problem in statistical physics which  in a much broader context  aims at determining  the response of a medium to a perturbation created by a driven TP  \cite{Marconi:2008,Cugliandolo:2011,vulp}. A considerable amount of knowledge has been gathered on the  forms of the so-called force-velocity relation, that being, the dependence of the TP velocity $v$ on the value of the applied force $F$, both in the linear and the non-linear response regimes.

 
Behavior beyond the force-velocity relation was  recently addressed numerically in the pioneering work \cite{Winter:2012},  which studied via Molecular Dynamics simulations the dynamics of an externally driven, or biased TP in a glass-forming liquid (a dense binary Yukawa liquid).
It was recognized that whilst the TP moves ballistically, i.e., $\langle X_t\rangle \sim v t$, the variance $\sigma_x^2 = \left<(X_t - \left<X_t\right>)^2\right>$ of the  TP position $X_t$ along the bias grows surprisingly in a superdiffusive manner with respect to time $t$, so that $\sigma_x^2 \sim t^{\lambda}$, where $\lambda$ is within the range  $1.3-1.5$. For such systems, this effect was found {\it only} in the close vicinity of the glass transition while regular diffusion was recovered away from the transition \cite{Harrer:2012}, suggesting that such anomalous behavior could be a distinct feature of being close to the glass transition.  

 
 A number of attempts  has been made to explain these findings,  based   either on a random trap model \cite{Winter:2012}, mode coupling theory \cite{Harrer:2012} or continuous-time random walks (CTRWs)  \cite{Schroer:2013}. All of these approaches rely on the notion  of a complex energy landscape and thereby assume that the  system is close to the glass transition. However, they  do not provide a quantitative nor qualitative understanding of the superdiffusive behaviour. In particular, the question  whether superdiffusion  is the ultimate regime or only a transient is still open \cite{Winter:2013,Schroer:2013}. 
   
Here, we show that  in fact superdiffusion in active microrheology settings can appear away from the glass transition, and even {\it independently} of glassy properties.  Based on a simple model that does not involve any complex energy landscape or kinetic constraints, we demonstrate that superdiffusion emerges  universally in  confined crowded systems.  We  fully quantify this superdiffusion, show that it is long-lived, highlight the key role of the system's geometry, and provide a clear physical mechanism underlying such behavior.

 
Our starting point is  a minimal model of a crowded system in which the motion of particles and of  the TP is mediated by so-called "diffusive" packing defects, or vacancies that are sufficiently large in size to allow their diffusive motion to proceed through direct swapping of their positions with the host medium particles. 
Note that the existence of such defects is tacitly assumed in various models of crowded systems \cite{Kurchan:1997,Ritort:2003,hoefling2013}.
Within this picture, the TP is released from the cage when a defect arrives to its location.

\begin{figure}\begin{center}
\includegraphics[width=6cm]{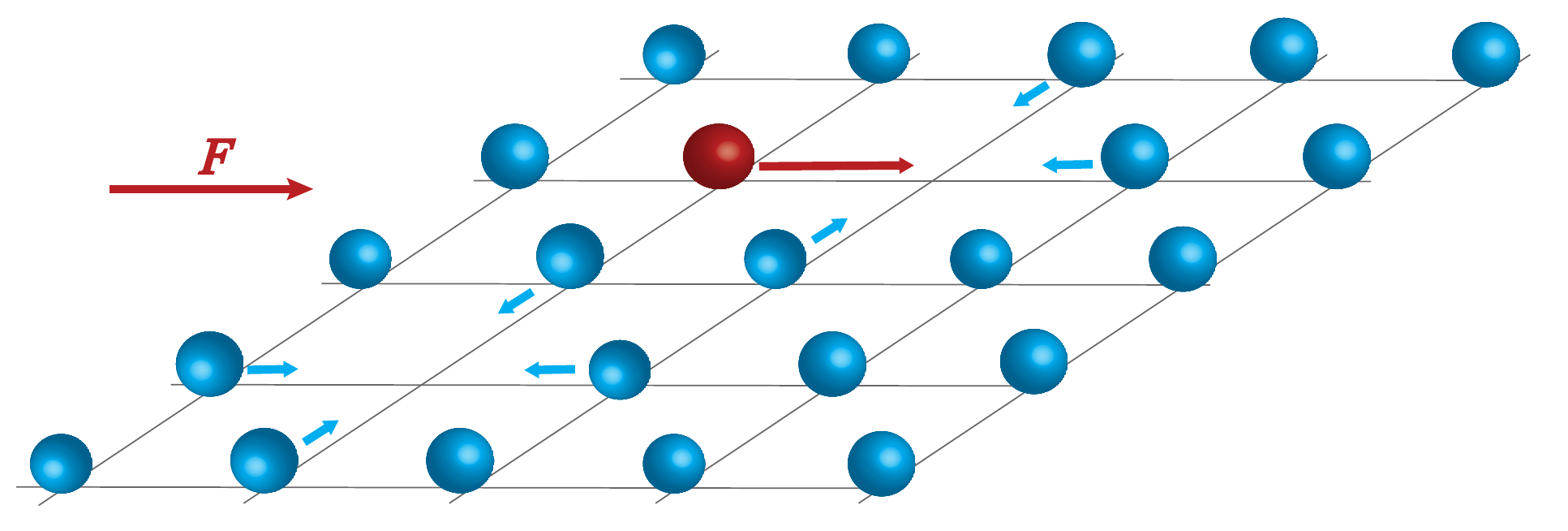}\end{center}
\caption{The model; a discrete lattice in which sites are occupied by identical hard-core medium particles (blue spheres). The red sphere denotes the tracer particle (TP) which, in addition to hard-core interactions,  is subject to an external force ${\bf F}=F{\bf e}_1$, and thus has asymmetric hopping probabilities. The arrows of different size depict schematically the hopping probabilities; a larger arrow near the TP indicates that it has a preference for moving in the direction of the applied field. Jumps are possible only when a vacancy (in concentration $\rho_0$) is adjacent to a particle. }
\label{model}
\end{figure}

More precisely, we consider a lattice gas model where the particles in the medium perform symmetric random walks on a $d$-dimensional lattice constrained by hard-core interactions between the particles, so that there is at most one particle per lattice site. The TP performs a random walk biased by an external applied force ${\bf F}= F {\bf e}_1$ (see figure \ref{model} and the SI for a detailed description of the dynamics), hindered by hard-core interactions with the host particles.
Note that this description represents a combination of two paradigmatic models of non-equilibrium statistical mechanics: asymmetric (for the TP) and symmetric (for the host particles)  simple exclusion processes \cite{Chou:2011}. Until now, only the force-velocity relation has been analyzed for such a   model \cite{Benichou:2000gd,Benichou:2001a}, with the exception of single file systems for which the variance has been calculated \cite{Benichou:2013a,Illien:2013} and infinite 2D systems, where however only a particular limit of the variance was considered \cite{Benichou:2002qq,Benichou:2013}. 


As noticed  in \cite{Brummelhuis:1989a} for two-dimensional systems, an important technical point is that in the absence of driving force and for small values of the density of vacancies $\rho_0$, the dynamics of the TP  can be deduced from analysing the joint dynamics of the TP and a single isolated vacancy.  As a matter of fact, this is still true for a biased TP in any dimension. In the SI, we show how to account for both the non trivial waiting time distribution and the anti-correlation effects between consecutive steps of the driven tracer, which are in fact induced by  the dynamics of a  single diffusing vacancy.
Exact asymptotic expressions of the variance $\sigma_x^2$ are obtained   for various geometries and for arbitrary values of the dimensionless force $f = \beta \sigma F$ (where $\sigma$ is the lattice step and $\beta$ is the reciprocal temperature). These are valid at large times and  low vacancy densities, and are  summarized below.

\begin{figure}\begin{center}
\includegraphics[width=9cm]{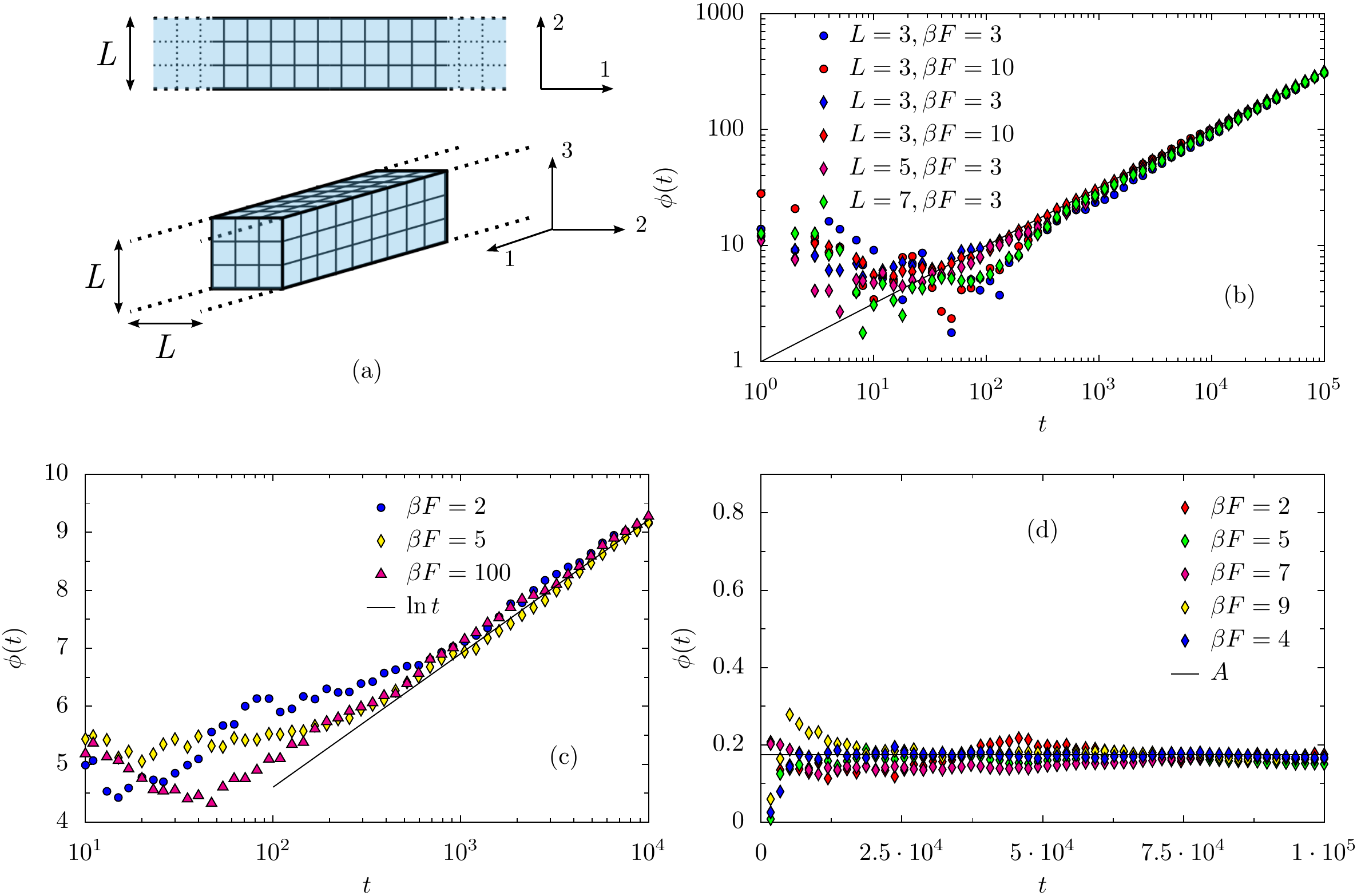}\end{center}
\caption{Studied geometries and reduced variance as a function of time in the superdiffusion regime. (a) : Sketch of stripe and capillary-like geometries. (b) : Simulations in capillaries (circles, $\phi(t)= \sqrt{3\pi/2} L^2/(4a_0^2\rho_0 t) \sigma_x^2(t) $) and stripes (diamonds, $\phi(t) = 3\sqrt{\pi}L/(8a_0^2\rho_0 t) \sigma_x^2(t)$) with density $\rho_0 = 10^{-5}$, and theoretical prediction (solid line : $\sqrt{t}$). (c) : Simulations on a 2D lattice with density $\rho_0=10^{-5}$ and $\phi(t)=\frac{\pi}{2 a_0}\left(\frac{\sigma_x^2(t)}{\rho_0 a_0 t} - \frac{2 a_0}{\pi}(\ln 8+\gamma -1) - 2a_0 \frac{\pi(5-2\pi)}{2\pi-4} -\coth( f/2) \right)$ and theoretical prediction (solid line : $\ln t$). (d) : Simulations on a 3D lattice with density $\rho_0 = 10^{-6}$ and $\phi(t) = [\sigma_x^2(t)/\rho_0t -a_0\coth( f/2)]/2a_0^2$ and theoretical prediction (solid line : $A$, defined after  Eq. (\ref{general})).}
\label{superdiff}
\end{figure}

{\it Superdiffusive regime.} First, our approach predicts  the following large-$t$ behaviour of the variance $\sigma_x^2$ in the leading order of $\rho_0$:
\begin{equation}
\label{general}
\lim_{\rho_0\to 0}\frac{\sigma_x^2}{\rho_0} \equi{t\to\infty} 2 a_0^2 t \times
\begin{cases}
 (4/3\sqrt{\pi} L) \, t^{1/2} & \text{2D stripe},\\
  (2\sqrt{2/3 \pi}/ L^2) \, t^{1/2} & \text{3D capillary},\\
\pi^{-1} \ln(t)& \text{2D  lattice},\\
 A+\coth(f/2)/(2a_0)& \text{3D   lattice,}
\end{cases}
\end{equation}
where $a_0$ is an $f$-dependent constant
\begin{eqnarray}
a_0 & =&  \frac{\sinh( f/2)}{\cosh( f /2)\left[1+\frac{2d\alpha}{2d-\alpha}\right]+d-1},
\end{eqnarray}
$A={\widehat P}(\mathbf{0}|\mathbf{0} ;1)+2(13\alpha-6)/[(2+\alpha)(\alpha-6)]$,
$d$ is the system dimension,
$\alpha  = \lim_{\xi\rightarrow 1^-}[{\widehat P}(\mathbf{0}|\mathbf{0};\xi)-{\widehat P}(2\mathbf{e_1}|\mathbf{0};\xi)]$ and
${\widehat P}(\mathbf{r}|\mathbf{r_0};\xi)$ is the generating function (discrete Laplace transform) of the propagator of a symmetric simple random walk (see the SI for the explicit expressions).
These surprisingly simple exact expressions unveil the dependence of the variance on time, width $L$ of the stripe or of the capillary, and on the reduced driving force $f$. Figure \ref{superdiff} shows an excellent quantitative agreement between the analytical predictions and the numerical simulations: The time, width and driving force dependences are unambiguously captured by our theoretical expressions.

A number of important conclusions can be drawn from this result: (i) Strong superdiffusion with exponent $\lambda =3/2$ takes place in confined, quasi-1D geometries, those being, infinitely long 3D capillaries and 2D stripes. This result is  quite  counterintuitive: indeed, in the absence of driving force it is common to encounter diffusive, or even subdiffusive growth of the fluctuations of the TP position in such crowded molecular environments \cite{hoefling2013}, however not superdiffusion
; (ii)  the superdiffusion in such systems emerges beyond (and therefore can not be reproduced within) the linear response-based approaches: The prefactor in the superdiffusive law is proportional to $f^2$ when $f \to 0$.  Despite the presence of the superdiffusion, the Einstein relation is nonetheless valid for systems of arbitrary geometry due to subdominant (in time) terms whose prefactor is proportional to $f$; (iii) in unbounded 3D systems $\sigma_x^2$ grows diffusively and not superdiffusively; (iv)  for $d=1$ (single files), one finds $\alpha =2$, so that $a_0=0$, and no super-diffusion can take place, in agreement with \cite{Illien:2013}. As a matter of fact, in this case the variance grows sub-diffusively.
Finally, this shows that superdiffusion is geometry-induced and the recurrence of the random walk performed by a vacancy is a necessary but not sufficient condition  in order for superdiffusion to occur.

{\it Giant diffusion regime.} The exact analytical result in Eq.(\ref{general}) 
provides explicit criteria for superdiffusion to occur. Technically, this yields the behavior of the variance  when the limit $\rho_0 \to 0$ is taken before the large-$t$ limit. It  however does not allow us, due to the nature of the limits involved, to answer the question whether the superdiffusion  is the ultimate regime (or just a transient), which requires the determination of  $\lim_{t\to\infty}\sigma_x^2$ at fixed $\rho_0$.  Importantly, we find that the order in which these limits are taken is crucial  in confined geometries ($\lim_{t\to\infty}\lim_{\rho_0\to0} \sigma_x^2 \neq \lim_{\rho_0\to0} \lim_{t\to\infty}\sigma_x^2$). In fact, the effective  bias experienced by a vacancy between two consecutive interactions with the TP, originating from  a non zero velocity of the TP, dramatically affects the ultimate long-time behavior of the variance in confined geometries.

More precisely, we show (see the SI) that the superdiffusive regime is always transient for an experimentally relevant system with $\rho_0$ fixed, while the long-time behaviour obeys
\begin{equation}\label{largetime}
\lim_{t\to\infty}\frac{\sigma_x^2}{t} \equi{\rho_0 \to 0}
\begin{cases}
B & \text{quasi-1D},\\
4 a_0^2\pi^{-1} \rho_0 \ln(\rho_0^{-1}) & \text{2D lattice},\\
2a_0^2[A+ \coth(f/2)/(2a_0)]\rho_0  & \text{3D lattice},
\end{cases}
\end{equation}
i.e., is always diffusive.  The constant $B$ depends on the driving force $f$ and on the geometry (quasi-1D in this case) of the system (see the SI). This long-time diffusive behaviour has several remarkable features: In dense quasi-1D systems the variance is independent of $\rho_0$, meaning that the corresponding longitudinal diffusion coefficient $D_{\parallel}$ is enhanced in comparison to the transverse one $D_{\perp}$ by a factor $1/\rho_0$, which may attain giant values in systems with $\rho_0 \ll 1$.
In 2D this effect is negligible and $D_{\parallel}$ is only a factor $\ln(\rho_0^{-1})$ larger than  $D_{\perp}$. In unbounded 3D systems no such strong anisotropy between $D_{\parallel}$ and  $D_{\perp}$ will emerge.

\begin{figure}\begin{center}
\includegraphics[width=8cm]{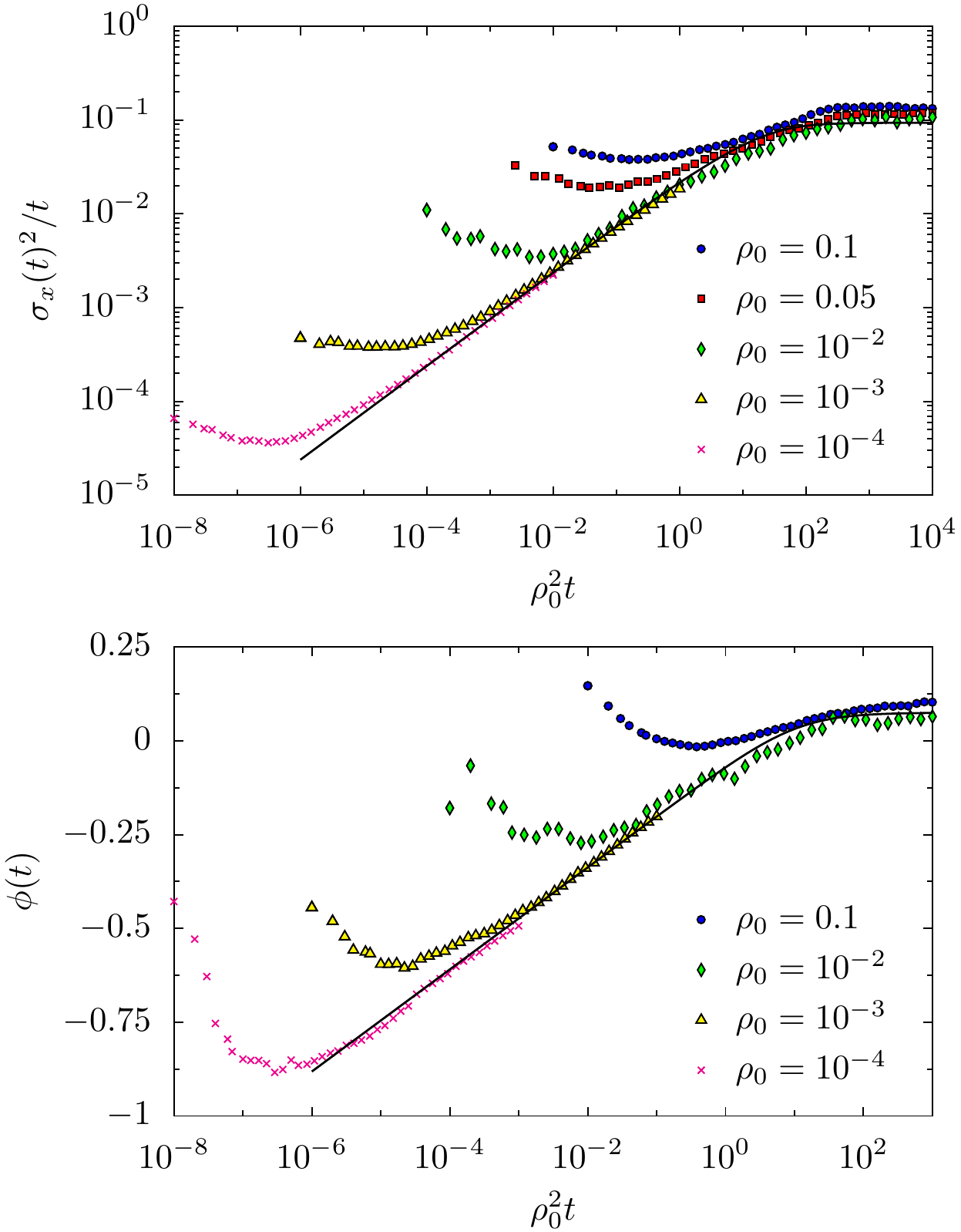}\end{center}
\caption{Top : Rescaled variance as a function of rescaled time $\rho_0^2 t$ on stripe-like lattices ($L=3$) for different densities ; solid line : $g(\rho_0^2 t)$, see the SI. Bottom : Rescaled variance $\phi(t) = \frac{\sigma_x^2(t)}{\rho_0 t} - \frac{2 a_0^2}{\pi} \ln \frac{1}{\rho_0^2 a_0^2}$ as a function of rescaled time $\rho_0^2 t$ on a 2D infinite lattice for different densities ; solid line : $=h(\rho_0^2 t)$ with $h(x)=\frac{2 a_0^2}{\pi} \ln \frac{a_0^2 x}{1+a_0^2 x}  + a_0\coth( f/2) + 2a_0^2  \frac{\pi(5-2\pi)}{2\pi-4} + \frac{2a_0^2}{\pi}\left( \ln 8 +\gamma-1   \right)$.}
\label{time}
\end{figure}

{\it Full dynamics: scaling regime and cross-over.} Finally, our approach provides the complete time evolution of the variance in the regime corresponding to $\rho_0\ll1$ and at a sufficiently large time $t$, that interpolates between the two limiting regimes of superdiffusion and giant diffusion listed above. In this regime, it is found that
\begin{equation}
\label{allregimes}
\sigma_x^2 \sim
\begin{cases}
t g(\rho_0^2 t)   & \text{quasi-1D},\\
- \frac{2 a_0^2}{\pi} \rho_0 t \ln \left((\rho_0 a_0)^2+1/t\right)  & \text{2D lattice},\\
2a_0^2 [A+ \coth(f/2)/(2a_0)] \rho_0  t& \text{3D lattice},
\end{cases}
\end{equation}
where the scaling function $g$ is explicitly given (see the SI) and satisfies $g(x)\equi{x\to0} x^{1/2}$ and $g(x)\equi{x\to\infty}{\rm const}$. Figure (\ref{time})
reveals excellent quantitative agreement between the analytical predictions and the numerical simulations. Several comments are in order: (i) Equation (\ref{allregimes}) encompasses both limiting behaviors (\ref{general}) and (\ref{largetime}), and shows explicitly 
that the 
limits $t\to\infty$ and $\rho_0\to0$ in quasi-1D and 2D systems do not commute; (ii)  In such systems the superdiffusion persists up until the onset of the giant diffusive behaviour at times $t_\times \sim 1/\rho_0^2$, which can be very large when $\rho_0 \ll 1$. Superdiffusion is therefore  very long-lived in such systems. Despite its transient feature, we thus expect superdiffusion to be a robust characteristic of   confined  crowded systems and we  anticipate that the ultimate diffusive behavior might be in practice difficult to observe.


The physical mechanism responsible for the emergence of the geometry-induced superdiffusion, revealed by our exact approach, can be ascertained in the large $f$ limit by a mean-field version of the model, which stipulates that after each interaction between the TP and a vacancy, all the other vacancies remain uniformly distributed. The model can then be reformulated as an effective CTRW  that takes into account {\it explicitly} the dynamics of the diffusive vacancies. This is in contrast to
the CTRW approach presented in \cite{Schroer:2013} for glassy systems, which 
infers the mean and the variance of the waiting-time distribution from the numerical data.
In the quasi-1D case, the waiting time of the first jump of the TP is extracted from the distribution $-dS_1/d\tau$, where $S_1$ is the well-known survival probability of an immobile target in a sea of diffusing predators $S_1(\tau)\propto \exp(-\rho_0\sqrt{\tau})$ \cite{Redner:2001a}. Waiting times of subsequent jumps are then drawn from the distribution $-d(T(\tau)S_1(\tau))/d\tau$, where $T(\tau)$ is the survival probability of an immobile target chased by a single random walker that starts near the target \cite{Redner:2001a}. 
Using the waiting time distribution described here, standard calculations (described in the SI) show that:  
(i) Super-diffusion with an exponent $\lambda=3/2$ appears as a result of repeated interactions between the TP and a {\it single}  vacancy in quasi-1D systems. This explains, in particular, why no superdiffusion takes place in strictly single-file systems, for which the cumulative displacement of the TP due to interactions with  a single vacancy amounts to at most one lattice step;  (ii)  diffusive behavior   is established  ultimately, when other vacancies start to interact with the  TP,  after a cross-over time which scales as $1/\rho_0^2$.
Note that while this mean-field approach reproduces the scaling properties of the variance 
with respect to the time and the density, it is unable to predict the correct width and driving force dependencies provided by our exact treatment.
 

 \begin{figure}[!h]\begin{center}
\includegraphics[width=8cm]{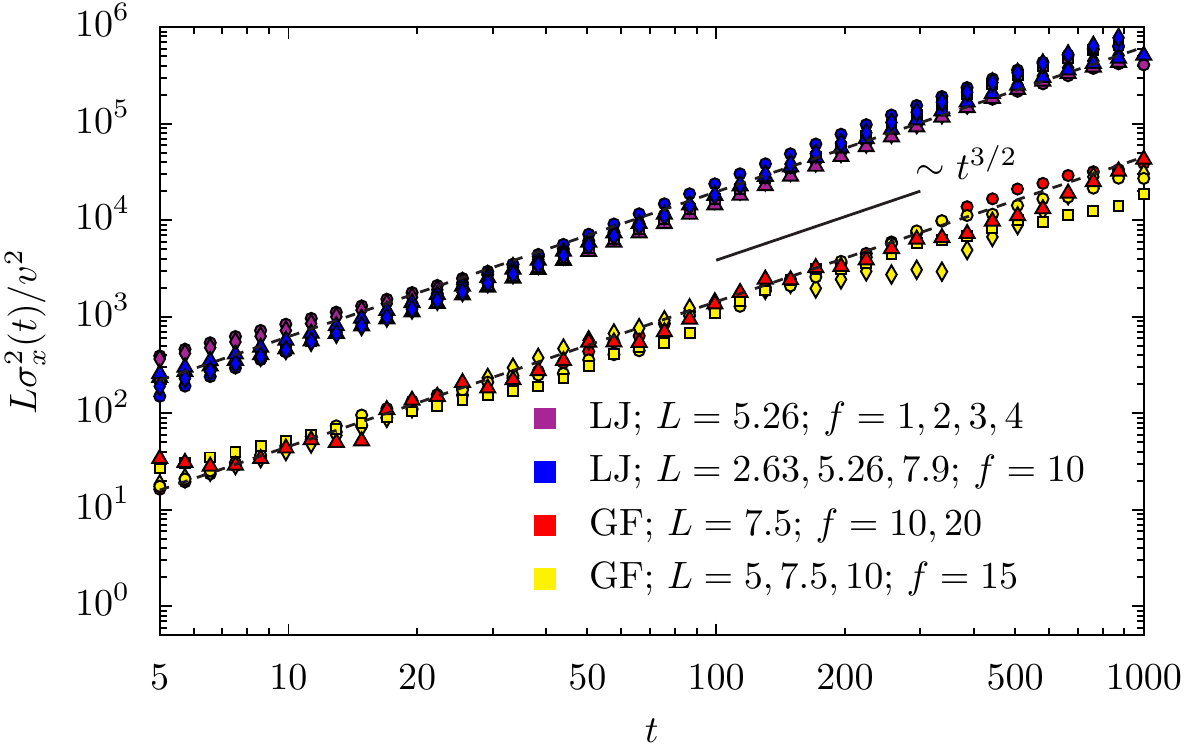}\end{center}
\caption{Rescaled variance $L^2 \sigma_x^2(t)/v^2$ as a function of time obtained from off-lattice simulations for different widths of stripes $L$ and forces $f$ (in the SI, it is shown that $v\sim a_0$ in the super diffusion  regime).  LJ : molecular dynamics of particles interacting via a Lennard-Jones potential in confined strip-like geometries. GF : simulations of  dense monodisperse granular fluid in confined strip-like geometries. More details on off-lattice simulations are given in the SI.}
\label{off}
\end{figure}

Altogether, our results show that the emergence of superdiffusion of a driven TP  crucially depends on the system's geometry, an aspect so far disregarded in this context. In order to quantitatively confirm our predictions on further crowded non-glassy systems, we performed   off-lattice  simulations investigating the dynamics of a biased TP confined to a controlled quasi-1D geometry  for models of monodisperse dense liquids (Lennard-Jones particles) and monodisperse granular fluids (using the algorithm presented in \cite{Fiege:2012}). 
In Fig. \ref{off} we plot the properly rescaled variance, where a clear data collapse is visible.  This   validates the time, width and driving force dependences that feature in our analytical expression (\ref{general}) also for off-lattice systems. Finally, our analysis shows that superdiffusion   is not the hallmark of glassy systems but is a generic feature of driven dynamics in confined crowded systems.

\vspace{1cm}

{\bf Acknowledgment} Support from European Research Council starting Grant No. FPTOpt-277998,  the Academy of Finland and  the EU IRSES DCP-PhysBio N269139 project is acknowledged.


\onecolumngrid

\newpage

\begin{center}

\Large{\textbf{Geometry-induced superdiffusion in driven crowded systems}}

\Large{\textbf{Supplementary information}}

\end{center}

\section{Superdiffusive regime}

\subsection{Model and notations}
\label{Model}

We consider a hypercubic lattice in dimension $d$, which can be infinite in every directions or partially finite with periodic boundary conditions (for example, we will consider a two-dimensional strip or a three dimensional capillary, equivalent to quasi 1D systems). 

 The lattice sites are occupied by hard-core particles performing symmetrical random walks, with the restriction that the occupancy number for each site is at most equal to one. The vacancy density is denoted as $\rho_0$. A tracer particle (denoted as TP in what follows) is also present on the lattice, and performs a biased random walk. In the usual fashion, the jump probability of the TP in direction $\nu$ is given by 
\begin{equation}
\label{ }
p_\nu = \exp\left( \frac{1}{2}\beta\sigma\mathbf{F}\cdot {\bf e}_{\boldsymbol \nu} \right) \left/  \sum_{\mu\in\{\pm 1,... \pm d\} }  \exp\left( \frac{1}{2}\beta\sigma\mathbf{F}\cdot {\bf e}_{\boldsymbol \mu} \right)\right.
\end{equation}
where $\beta$ is the inverse temperature, $\sigma$ the characteristic length of a single move and $\mathbf{F}$ a drag force. For simplicity and with no loss of generality we will assume that $\mathbf{F} = F \mathbf{e_1}$. In what follows, we introduce the dimensionless force $f=\beta \sigma F$. The probability for the TP to be on site $\mathbf{X}$ at time step $t$ is denoted $P_t(\mathbf{X})$. The random variable representing the first component of $\mathbf{X}$ at time $t$ is $X_t$. 

At each time step $t$, a vacancy is randomly selected and moves according to the following rules :
\begin{itemize}
  \item if the TP is not adjacent to the vacancy, one neighbor is randomly selected and exchanges its position with the vacancy.
  \item otherwise, if the vacancy is at position $\mathbf{Y}$ and the TP at position ${\bf Y}+{\bf e}_{\boldsymbol \nu}$, the TP exchanges its position with the vacancy with probability $p_\nu/[(1-1/2d)+p_\nu]$ and with probability $1/(2d-1+2d p_\nu)$ with any of the $2d-1$ other neighbor.
  \item if one of the neighboring sites is a vacancy, the latter is treated as an unbiased walker.
\end{itemize}

In what follows, we will focus on the expression of the propagator in the large-time and low-density limits, and will show that these two limits do not commute.

\subsection{Single-vacancy problem}

We first solve the single-vacancy problem. There is only one vacancy on the lattice, whose position is denoted by $\mathbf{Y}$ (its initial position is $\mathbf{Y_0}$). The TP is initially at the origin of the lattice. We introduce the following notations :
\begin{itemize}
	\item $P^{(1)}_t(\mathbf{X}|{\bf Y_0})$ is the probability that the TP is at site $\mathbf{X}$ at time moment $t$ given that the vacancy was initially at site $\mathbf{Y_0}$.
	\item $F^*_t({\bf 0}|{\bf Y_0})$ is the probability that the
vacancy, which starts its random walk at the site ${\bf Y}_0$,
 arrives at the origin  ${\bf 0}$ for the first time at the time step $t$.
	\item $F^*_t({\bf 0}|{\bf e}_{\boldsymbol \nu}|{\bf Y_0})$ is the
conditional probability that the vacancy, which starts its random walk at the site ${\bf Y}_0$, 
 appears  at the origin for the first
time at the time step $t$, being at time moment $t - 1$ at the site 
${\bf e}_{\boldsymbol \nu}$. 
\end{itemize}

With these notations, we get 
\begin{eqnarray}
P^{(1)}_t({\bf X}|{\bf Y_0})&=& \delta_{{\bf X},{\bf 0}}\left(1-\sum_{j=0}^t F^*_j({\bf
0}|{\bf Y_0})\right) +\sum_{p=1}^{+\infty}\sum_{m_1=1}^{+\infty}\ldots\sum_{m_p=1}^{+\infty}
\sum_{m_{p+1}=0}^{+\infty}\delta_{m_1+\ldots+m_{p+1},t}\sum_{\nu_1}\ldots\sum_{\nu_p}\delta_{{\bf
e}_{{\boldsymbol \nu}_1}+\ldots+{\bf
e}_{{\boldsymbol \nu}_p},{\bf X}}  \nonumber\\
&\times& \left(1-\sum_{j=0}^{m_{p+1}}F^*_j({\bf
0}|-{\bf e}_{{\boldsymbol \nu}_{\bf p}})\right)  F^*_{m_p}({\bf 0}|{\bf e}_{{\boldsymbol \nu}_{\bf p}}|-{\bf
e}_{{\boldsymbol \nu}_{\bf p-1}})\ldots F^*_{m_2}({\bf 0}|{\bf e}_{{\boldsymbol
\nu}_{\bf 2}}|-{\bf e}_{{\boldsymbol \nu}_{\bf 1}}) F^*_{m_1}({\bf 0}|{\bf
e}_{{\boldsymbol \nu}_{\bf 1}}|{\bf Y_0}).
\label{Ptr}
\end{eqnarray}
Denoting by $\widehat{\cdot}$ the discrete Laplace transform (generating function) and by $\widetilde{\cdot}$ the Fourier transform, we get
\begin{equation}
\label{P1def}
\widehat{\widetilde{P}}^{(1)}({\bf k}|{\bf Y_0};\xi)=\frac{1}{1-\xi}\left(1+{\cal D}^{-1}({\bf
k};\xi)\sum_{\mu}U_{ \mu}({\bf k};\xi)F^*({\bf
0}|{\bf e}_{\boldsymbol \mu}|{\bf Y_0};\xi)\right)
.
\end{equation}
In Eq. (\ref{P1def}) 
the function ${\cal D}({\bf k};\xi)$ stands for 
the determinant of the following $2d \times 2d$ matrix,
\begin{equation}
\label{Ddef}
{\cal D}({\bf k};\xi)\equiv{\rm det}({\rm{\bf  I-T}}({\bf k};\xi)),
\end{equation} 
where $\mathbf{I}$ is the identity of size $2d$ and the matrix ${\rm {\bf T}}({\bf k};\xi)$ has the elements $\left({\rm {\bf T}}({\bf k};\xi)\right)_{\nu,\mu}$  defined by
\begin{equation}
\label{Tdef}
\left({\rm {\bf T}}({\bf k};\xi)\right)_{\nu,\mu} = \exp[i ({\bf k} \cdot {\bf
e_{\boldsymbol \nu}})]  A_{\nu,-\mu}(\xi).
\end{equation}
The coefficients $A_{\nu,\mu}(\xi)$ ($\nu,\mu = \pm 1, ..., \pm d$),  stand for
\begin{equation}
A_{\nu,\mu}(\xi) \equiv F^*({\bf 0}|{\bf e}_{\boldsymbol \nu}|{\bf e}_{\boldsymbol \mu} ; \xi) = \sum_{t = 0}^{+ \infty}  
F^*_t({\bf 0}|{\bf e}_{ \boldsymbol \nu}|{\bf e}_{\boldsymbol \mu}) \xi^t.
\end{equation}
Lastly, the matrix $U_{{ \mu}}({\bf k};\xi)$ in Eq. (\ref{P1def}) is given by
\begin{eqnarray}
\label{Udef}
U_{\mu}({\bf k};\xi)\equiv{\cal D}({\bf k};\xi)\sum_{ \nu} 
\left[1-\ex{-i {\bf k} \cdot {\bf e}_{\boldsymbol \nu}}\right]({\bf I} -{\bf T}({\bf
k};\xi))^{-1}_{\nu,\mu}  \ex{i {\bf
k} \cdot {\bf e}_{\boldsymbol \mu}}.
\end{eqnarray}

Consequently, $\widehat{\widetilde{P}}^{(1)}({\bf k}|{\bf Y_0};\xi)$ may be calculated if the functions $A_{\nu,\mu}(\xi)$ are explicitly known (see details below).\\



 Let $P_t(\mathbf{r}|\mathbf{r_0})$ be the propagator of an unbiased walker on the considered lattice (i.e. the probability for the walker to be at site $\mathbf{r}$ at time $t$ knowing that it started from site $\mathbf{r_0}$). The return probabilities $A_{\nu,\mu}(\xi)$ can then be expressed in terms of the generating function $\widehat{P}(\mathbf{r}|\mathbf{r_0} ;\xi)$ of $P_t(\mathbf{r}|\mathbf{r_0})$ by adding a defective, absorbing state at the origin \cite{PRE2002, Hughes}. Then, the vacancy random walk can be formally represented as a lattice random walk with site-dependent probabilities of the form
 $p^+({\bf s}|{\bf s'}) = 1/(2d)+q({\bf s}|{\bf s'})$, where ${\bf s'}$ is the site occupied by the vacancy at the time moment $t$,  
${\bf s}$ denotes the target, nearest-neighboring to ${\bf s'}$ site, 
\begin{equation}
q({\bf s}|{\bf s'})\equiv
\begin{cases}
0& \text{if ${\bf s'}\notin\{{\bf 0},{\bf e_{\pm 1}},\dots, {\bf e_{\pm d}}\}$},\\
\delta_{{\bf s},{\bf 0}}-1/(2d)& \text{if ${\bf s'}={\bf 0}$},\\
\delta q_\nu& \text{if ${\bf s'}={\bf e}_{\boldsymbol \nu}$ and ${\bf s}={\bf
0}$},\\
-\delta q_\nu/(2d-1)& \text{if ${\bf s'}={\bf e}_{\boldsymbol \nu}$ and ${\bf
s'}\neq{\bf 0}$},
\end{cases}
\end{equation} 
where $\delta q_\nu$ is defined by 
\begin{equation}
\delta q_\nu\equiv\frac{p_\nu}{p_\nu+\frac{2d-1}{2d}}-\frac{1}{2d}
\end{equation}
Further on, we define $P_{t}^+({\bf s}|{\bf s_0})$ as the probability
distribution associated with such a random walk starting at site ${\bf
s_0}$ at step $t = 0$. 

Now, let the symbols ${\cal E}$,  ${\cal A}$ and  ${\cal B}$ define the following three events:
\begin{itemize}
\item  the event ${\cal E}$: the vacancy, which has started its random walk at the site ${\bf e}_{\boldsymbol \nu}$, 
visits the
origin ${\bf 0}$ for the first time at the
$t$-th step exactly, being at  the site ${\bf e}_{\boldsymbol \mu}$ at the previous step $t-1$;
\item  the event ${\cal A}$: the vacancy, which started its random walk at the site  ${\bf e}_{\boldsymbol \nu}$,
is at the site ${\bf e}_{\boldsymbol \mu}$
at the time moment $t-1$ and the origin ${\bf 0}$ has not been visited during the $t-1$ first
steps of its walk;
\item  the event ${\cal B}$: the vacancy jumps from the neighboring to the origin 
site ${\bf e}_{\boldsymbol \mu}$
to the site ${\bf 0}$ at the $t$-th step exactly. 
\end{itemize}

Evidently,  by definition, the desired first visit probability $F^*_t({\bf 0}|{\bf e}_{\boldsymbol \mu}|{\bf e}_{\boldsymbol \nu})$
is just  the probability of the ${\cal E}$ event 
\begin{equation}
\label{99}
F^*_t({\bf 0}|{\bf e}_{\boldsymbol \mu}|{\bf e}_{\boldsymbol \nu}) = {\rm Prob}({\cal E}).
\end{equation}
To calculate ${\rm Prob}({\cal E})$ we note first that 
the probabilities of such three events obey:
\begin{equation}
\label{z}
{\rm Prob}({\cal E})={\rm Prob}({\cal A}\cap{\cal B})={\rm Prob}({\cal A}){\rm Prob}({\cal B}).
\end{equation}
On the other hand, we have that
\begin{equation}
\label{zz}
{\rm Prob}({\cal A})=P_{t-1}^+({\bf
e}_{\boldsymbol \mu}|{\bf e}_{\boldsymbol \nu}), 
\end{equation}
and
\begin{equation}
\label{zz1}
{\rm
Prob}({\cal B})=p^+(\emuu|\zz).
\end{equation} 
Hence, in virtue of Eqs.(\ref{99}),(\ref{z}),(\ref{zz}) and (\ref{zz1}), the return probability 
$F^*_t({\bf 0}|{\bf e}_{\boldsymbol \mu}|{\bf e}_{\boldsymbol \nu})$ is given explicitly by its Laplace transform
\begin{equation}
\widehat{F}^*({\bf 0}|{\bf e}_{\boldsymbol \mu}|{\bf e}_{\boldsymbol
\nu};\xi)=\xi p^+(\emuu|\zz) \widehat{P}^+({\bf
e}_{\boldsymbol \mu}|{\bf e}_{\boldsymbol \nu};\xi).
\end{equation}
Therefore, the calculation of the return
probabilities  $F^*_t({\bf 0}|{\bf
e}_{\boldsymbol \mu}|{\bf e}_{\boldsymbol \nu})$ amounts to the
evaluation of the probability distribution $P_{t}^+({\bf s}|{\bf s_0})$  of the vacancy random
 walk in the presence of an
absorbing site placed at the lattice origin.

Making use of the generating function technique adapted to random walks on lattices with defective sites \cite{Hughes}, we obtain
\begin{equation}
\widehat{P}^+({\bf s_i}|{\bf s_j};\xi)=\widehat{P}({\bf s_i}|{\bf s_j};\xi) +
\sum_{l=-d}^d \widehat{A}({\bf s_i}|{\bf s_l};\xi)\widehat{P}^+({\bf s_l}|{\bf
s_j};\xi),
\label{pasmatricielle}
\end{equation} 
where
\begin{equation}
{\bf s_i}\equiv
\begin{cases}
{\bf e_i},& \text{for $i\in\{\pm1,\dots, \pm d\}$},\\
{\bf 0},& \text{for $i=0$},
\end{cases}
\end{equation}
and
\begin{equation}
\widehat{A}({\bf s_i}|{\bf s_l};\xi)\equiv
\xi\sum_{{\bf s'}} \widehat{P}({\bf s_i}|{\bf s'};\xi)
q({\bf s'}|{\bf s_l}),
\end{equation}
$ \widehat{P}({\bf s_i}|{\bf s_j};\xi)$ being the generating
function of the unperturbed associated random walk (that is,
biased random walk 
with no defective sites).

Further on, Eq. (\ref{pasmatricielle}) can be recast into the following matricial form:
\begin{equation}
{\rm {\bf P^+}}=({\rm {\bf 1-A}})^{-1}{\rm {\bf P}},
\label{matricielle}
\end{equation} 
in which equation ${\rm {\bf P,P^+,A}}$ stand for the $(2d+1)\times(2d+1)$ matrices with the elements
defined by
\begin{equation}
\label{defineP}
{\rm {\bf P}}_{i,j}=\widehat{P}({\bf s_i}|{\bf s_j};\xi),{\rm {\bf P^+}}_{i,j}=
\widehat{P}^+({\bf s_i}|{\bf s_j};\xi),{\rm {\bf A}}_{i,j}=\widehat{A}({\bf s_i}|{\bf s_j};\xi),
\end{equation}
where $i,j = 0,\pm 1,\dots,\pm d$. One can readily express $\mathbf{A}$ and $\mathbf{P}$ (see below), and deduce $\mathbf{P}^+$ and the return probabilities $F^*$. 


%
%

\subsection{Finite low vacancy density}

We now start from a finite hypercubic lattice of $N$ sites with a finite number $M$ of vacancies. The vacancy density is $\rho_0=M/N$ and is supposed to be very small, $\rho_0 \ll 1$, so that we will restrain our analysis to the leading order in $\rho_0$. The TP is initially at the origin and initial positions of the vacancies are denoted by $\mathbf{Y}_\zz^{(1)}, \mathbf{Y}_\zz^{(2)}, \ldots, \mathbf{Y}_\zz^{(M)}$, which are all different from each other and from $\zz$. At each time step, all vacancies exchange their positions with one of the neighboring particles, such that each vacancy makes a jump each time step. The general expression of  $\mathcal{P}_t(\mathbf{X}|\mathbf{Y}_\zz^{(1)}, \mathbf{Y}_\zz^{(2)}, \ldots, \mathbf{Y}_\zz^{(M)})$, which denotes the probability of finding at time step $t$ the TP at position $\mathbf{X}$ as a result of its interactions with all $M$ vacancies is
\begin{equation}
\mathcal{P}_t(\mathbf{X}|\mathbf{Y}_\zz^{(1)}, \mathbf{Y}_\zz^{(2)}, \ldots, \mathbf{Y}_\zz^{(M)}) = \sum_{\mathbf{X}_\zz^{(1)}} \ldots \sum_{\mathbf{X}_\zz^{(M)}} \delta_{\mathbf{X},\mathbf{X}_\zz^{(1)}+\ldots+\mathbf{X}_\zz^{(M)}} \mathcal{P}_t(\mathbf{X}_\zz^{(1)}, \mathbf{X}_\zz^{(2)},\ldots, \mathbf{X}_\zz^{(M)}|\mathbf{Y}_\zz^{(1)}, \mathbf{Y}_\zz^{(2)},\ldots, \mathbf{Y}_\zz^{(M)})
\end{equation}
where $\mathcal{P}_t(\mathbf{X}_\zz^{(1)}, \mathbf{X}_\zz^{(2)},\ldots, \mathbf{X}_\zz^{(M)}|\mathbf{Y}_\zz^{(1)}, \mathbf{Y}_\zz^{(2)},\ldots, \mathbf{Y}_\zz^{(M)})$ denotes the conditional probability that within the time interval $t$ the TP has displaced to the position $\mathbf{X}_\zz^{(1)}$ due to its interactions with the first vacancy, to the position $\mathbf{X}_\zz^{(2)}$ due to its interactions with the second vacancy etc.

It may appear that two vacancies are adjacent or on the same site. Implementing the evolution of such situations would require additional dynamic rules, which are in fact unnecessary as such events would contribute only to order $\mathcal{O}(\rho_0^2)$ in the calculation. Consequently, at leading order in $\rho_0$, all vacancies contribute independently to the TP displacement and discarding the events involving two or more vacancies, it is possible to show \cite{PRE2002} that the propagator of the TP on the finite lattice can be expressed in term of the single-vacancy propagator in the following way
\begin{equation}
\widetilde{P}_t(\mathbf{k} ; M,N) \simeq \left[ \widetilde{P}^{(1)}_t(\mathbf{k}|{\bf Y_0}) \right]^M.
\end{equation}

Since we consider lattice geometries where the number of sites is infinite, we now turn to the thermodynamic limit $M, N \to \infty$ with fixed ratio $M/N = \rho_0$. Denoting as $\widetilde{P}_t({\bf k})$ the Fourier transform of the propagator of the TP with a fixed density, we get
\begin{equation}
\widetilde{P}_t(\mathbf{k} ) = \lim_{M, N\to \infty} \widetilde{P}_t(\mathbf{k} ; M,N) \simeq \exp[-\rho_0 \Omega_t({\bf k})]
\end{equation}
where
\begin{eqnarray}
\Omega_t({\bf k})  &  \equiv  &  \sum_{j=0}^t\sum_\nu\Delta_{t-j}({\bf k}|{\bf e}_{\boldsymbol
\nu})\sum_{{\bf Y\neq0}} F^*_j({\bf
0}|{\bf e}_{\boldsymbol \nu}|{\bf Y}), \\
\Delta_t({\bf k}|{\bf e}_{\boldsymbol \nu})  &=  &1-\widetilde{P}_t^{(1)}({\bf
k}|{\bf -e}_{\boldsymbol \nu}) \exp[i ({\bf k} \cdot {\bf e}_{\boldsymbol \nu})],
\end{eqnarray}
$ F^*_j({\bf
0}|{\bf e}_{\boldsymbol \nu}|{\bf Y})$ are the conditional return probabilities previously defined and $\widetilde{P}_t^{(1)}({\bf
k}|{\bf -e}_{\boldsymbol \nu})$ is the Fourier transformed single-vacancy probability distribution $P_t^{(1)}({\bf
X}|{\bf -e}_{\boldsymbol \nu})$. The latter can be readily obtained by applying the discrete Fourier transformation 
to the relation
\begin{equation}
P_t^{(1)}({\bf X}|{\bf Y})= \delta_{{\bf X},{\bf 0}}\left(1-\sum_{j=0}^t F^*_j({\bf
0}|{\bf Y})\right)+\sum_{j=0}^t\sum_\nu P_{t-j}^{(1)}({\bf X-e}_{\boldsymbol \nu}|{\bf -e}_{\boldsymbol\nu}) F^*_j({\bf
0}|{\bf e}_{\boldsymbol \nu}|{\bf Y}).
\end{equation} 
and chosing ${\bf Y } = {\bf -e}_{\boldsymbol \nu}$. 

Using Laplace-transformed quantities, we get the following expression for $\Omega$ :
\begin{equation}
\label{Omega}
\Omega({\bf k};\xi)=\sum_\nu\Delta({\bf k}|{\bf e}_{\boldsymbol \nu};\xi)\sum_{{\bf Y\neq0}} \widehat{F}^*({\bf
0}|{\bf e}_{\boldsymbol \nu}|{\bf Y};\xi),
\end{equation}
where
\begin{equation}
\label{ }
\Delta({\bf k}|{\bf e}_{\boldsymbol \nu};\xi)=\frac{1}{1-\xi}-\widehat{\widetilde{P}}^{(1)}({\bf k}|{\bf e}_{\boldsymbol \nu};\xi)\ex{i\mathbf{ k}\cdot\mathbf{ e}_{\boldsymbol \nu}}
\end{equation}

We now apply these general results to particular geometries, and focus on the behavior of the variance of the TP position $\sigma_x^2$, whose Laplace transform is linked to the $\Omega({\bf k};\xi)$ function throughout the relation :
\begin{equation}
\label{ }
\widehat{\sigma}_x^2(\xi) = \rho_0 \left. \frac{\partial^2 \Omega}{\partial k_1^2}\right|_{\mathbf{k}=\zz}
\end{equation}

Expressing the propagators $\widehat{P}({\bf s_i}|{\bf s_j};\xi)$ defined earlier in specific geometries, it is then straightforward to obtain an expression for the variance in the long-time ($\xi \rightarrow 1^-$) and low vacancy density ($\rho_0 \rightarrow 0$). Note that in this calculation, we first considered the $\rho_0 \rightarrow 0$ limit and then the large time limit. We present several explicit expressions in what follows. 

\subsection{Explicit expressions for particular geometries}

\subsubsection{Two-dimensional strip}
\label{superdiff_strip}

The propagators of a symmetric random walk on a two-dimensional strip ($L\times \infty$) can be written as
\begin{equation}
\widehat{P}(\mathbf{r}|\zz;\xi)  =\frac{1}{L}\sum_{k=0}^{L-1} \frac{1}{2\pi} \int_{-\pi}^\pi \dd q \frac{\ex{-i n_1 q} \ex{-2i\pi k n_2/L}}{1-\frac{\xi}{2} \left[\cos q +\cos \frac{2\pi k}{L}\right]}.
\end{equation}
and can be explicitly developed in the $\xi \rightarrow 1^-$ limit \cite{Hughes, Hilhorst} :
\begin{eqnarray}
g_1(\xi) & \equiv &   \widehat{P}(\mathbf{ 0}|\mathbf{ 0};\xi)    =  \frac{1}{L\sqrt{1-\xi}}+S_{L,-1}+\mathcal{O}(1-\xi) \label{g1}  \\
g_2(\xi) & \equiv &\widehat{P}(\mathbf{ e}_1|\mathbf{ 0};\xi)    =   \frac{1}{L\sqrt{1-\xi}}+S_{L,-1}+2S_{L,1}-2+\frac{2}{L}\sqrt{1-\xi}+\mathcal{O}(1-\xi) \\
g_3(\xi) & \equiv &\widehat{P}(\mathbf{ e}_2|\mathbf{ 0};\xi)  =   \frac{1}{L\sqrt{1-\xi}}+S_{L,-1}-2S_{L,1}+\mathcal{O}(1-\xi) \\
g_4(\xi) & \equiv &\widehat{P}(2\mathbf{ e}_1|\mathbf{ 0};\xi)   =   \frac{1}{L\sqrt{1-\xi}}+S_{L,-1}+8S_{L,1}+8S_{L,3}-8+\frac{8}{L}\sqrt{1-\xi}+\mathcal{O}(1-\xi) \\
g_5(\xi) & \equiv &\widehat{P}(2\mathbf{ e}_2|\mathbf{ 0};\xi)   =   \frac{1}{L\sqrt{1-\xi}}+S_{L,-1}-8S_{L,1}+8S_{L,3}+\mathcal{O}(1-\xi) \\
g_6(\xi) & \equiv &\widehat{P}(\mathbf{ e}_1+\mathbf{ e}_2|\mathbf{ 0};\xi)    =   \frac{1}{L\sqrt{1-\xi}}+S_{L,-1}-4S_{L,3}+\frac{2}{L}\sqrt{1-\xi}+\mathcal{O}(1-\xi)    \label{g6}  \\
\end{eqnarray}
where we defined the sums
\begin{equation}
\label{ }
S_{L,n}\equiv \frac{1}{L}\sum_{k=1}^{L-1}\frac{\sin^n(\pi k/L)}{\sqrt{1+\sin^2(\pi k/L)}}.
\end{equation}

The $5\times 5$ matrices $\mathbf{P}$ and $\mathbf{A}$ are then defined by

\begin{eqnarray}
\mathbf{P}&=&
\begin{pmatrix}
  g_1 & g_2 & g_3 & g_3 & g_3 \\
  g_2 & g_1 & g_4 & g_6 & g_6 \\
  g_2 & g_4 & g_1 & g_6 & g_6 \\
  g_3 & g_6 & g_6 & g_1 & g_5 \\
  g_3 & g_6 & g_6 & g_5 & g_1  \nonumber
  \end{pmatrix}\\ 
  \mathbf{A}&=&
 \begin {pmatrix}
     1- \left( 1-\xi \right) g_1&\frac{4}{3}
\xi q_{{1}} \left( g_1-{\frac {g_2}{\xi}} \right) &\frac{4}{3}\xi
q_{{-1}} \left( g_1-{\frac {g_2}{\xi}} \right) &\frac{4}{3}\xi
q_{{2}} \left( g_1-{\frac {g_3}{\xi}} \right) &\frac{4}{3}\xi q_{
{2}} \left( g_1-{\frac {g_3}{\xi}} \right) \\ \noalign{\medskip}
- \left( 1-\xi \right) g_2&\frac{4}{3}q_{{1}}
 \left( \xi g_2-g_1+1 \right) &\frac{4}{3}q_{{-1}} \left( \xi g_2-g_4 \right) &\frac{4}{3}q_{{2}} \left( \xi g_2-g_6
 \right) &\frac{4}{3}q_{{2}} \left( \xi g_2-g_6 \right) 
\\ \noalign{\medskip}- \left( 1-\xi \right) g_2&\frac{4}{3}q_{{1}}
 \left( \xi g_2-g_4 \right) &\frac{4}{3}q_{{-1}} \left( \xi g_2-g_1+1 \right) &\frac{4}{3}q_{{2}} \left( \xi g_2-g_6
 \right) &\frac{4}{3}q_{{2}} \left( \xi g_2-g_6 \right) 
\\ \noalign{\medskip}- \left( 1-\xi \right) g_3&\frac{4}{3}q_{{1}}
 \left( \xi g_3-g_6 \right) &\frac{4}{3}q_{{-1}} \left( \xi g_3-g_6 \right) &\frac{4}{3}q_{{2}} \left( \xi g_3-g_1+1
 \right) &\frac{4}{3}q_{{2}} \left( \xi g_3-g_5 \right) 
\\ \noalign{\medskip}- \left( 1-\xi \right) g_3&\frac{4}{3}q_{{1}}
 \left( \xi g_3-g_6 \right) &\frac{4}{3}q_{{-1}} \left( \xi g_3-g_6 \right) &\frac{4}{3}q_{{2}} \left( \xi g_3-g_5
 \right) &\frac{4}{3}q_{{2}} \left( \xi g_3-g_1+1 \right) \\ \nonumber
\end {pmatrix} ,
\end{eqnarray}

where we did not specify the $\xi$-dependency of the $g$ functions in order to simplify the notations. Using the definition of $\Omega$ from Eq. (\ref{Omega}), we finally obtain :
\begin{equation}
\label{varsuperdiffstrip}
\lim_{\rho_0\to 0}\frac{\sigma_x^2}{\rho_0} \underset{t \to \infty}{\sim}  \frac{8 a_0^2}{3\sqrt{\pi}L} t^{3/2}.
\end{equation}
where 
\begin{eqnarray}
a_0 &=&  \frac{\sinh( f/2)}{\cosh( f /2)\left[1+\frac{4\alpha}{4-\alpha}\right]+1} \\
\alpha & =& \lim_{\xi\rightarrow 1^-}[\widehat{P}(\mathbf{0}|\mathbf{0};\xi)-\widehat{P}(2\mathbf{e_1}|\mathbf{0};\xi)].
\end{eqnarray}

Using the following expression for the mean position of the TP
\begin{equation}
\label{ }
\moy{X}(\xi) = -i \rho_0 \left. \frac{\partial \Omega}{\partial k_1}\right|_{\mathbf{k}=\zz}
\end{equation}
it is also possible to show that
\begin{equation}
\lim_{\rho_0\to 0}\frac{\moy{X_t}}{\rho_0} \underset{t \to \infty}{\sim}  a_0 t.
\end{equation}


\subsubsection{Three-dimensional capillary}

Following the calculations presented by Hilhorst \cite{Hilhorst} for the strip geometry, we calculate the propagators of a symmetric P\'{o}lya-like random walk on a three-dimensional capillary ($L\times L \times \infty$). We get

\begin{eqnarray}
\label{ }
 \widehat{P}(\zz|\mathbf{ 0};\xi)  &=& \frac{\sqrt{6}}{2L^2\sqrt{1-\xi}}+S_0 +\mathcal{O}(1-\xi) \label{h1}\\
\widehat{P}(\mathbf{e_1}|\mathbf{ 0};\xi)  &=& \frac{\sqrt{6}}{2L^2\sqrt{1-\xi}}+3S_0-3-S_1+\frac{3\sqrt{6}}{2L^2}\sqrt{1-\xi} +\mathcal{O}(1-\xi)\\
\widehat{P}(\mathbf{e_2}|\mathbf{ 0};\xi) &=&  \frac{\sqrt{6}}{2L^2\sqrt{1-\xi}}+\frac{1}{2}S_1 +\mathcal{O}(1-\xi) \\
\widehat{P}(2\mathbf{e_1}|\mathbf{ 0};\xi)  &=& \frac{\sqrt{6}}{2L^2\sqrt{1-\xi}}-18+17S_0-12S_1+2S_2+\frac{6\sqrt{6}}{L^2}\sqrt{1-\xi} +\mathcal{O}(1-\xi)\\
\widehat{P}(2\mathbf{e_2}|\mathbf{ 0};\xi) &= &\frac{\sqrt{6}}{2L^2\sqrt{1-\xi}}+S_2-2T-S_0 +\mathcal{O}(1-\xi) \\
\widehat{P}(\mathbf{e_1}+\mathbf{e_2}|\mathbf{ 0};\xi) &=& \frac{\sqrt{6}}{2L^2\sqrt{1-\xi}} + \frac{3}{2}S_1-\frac{1}{2}S_2+\frac{3\sqrt{6}}{2L^2}\sqrt{1-\xi} +\mathcal{O}(1-\xi) \\
\widehat{P}(\mathbf{e_1}+\mathbf{e_2}|\mathbf{ 0};\xi) &=& \frac{\sqrt{6}}{2L^2\sqrt{1-\xi}} +T +\mathcal{O}(1-\xi) \label{h7}.
\end{eqnarray}

with

\begin{eqnarray}
S_n & = & \frac{1}{L^2}\sum_{(k_2,k_3)\neq(0,0)}^{L-1}\frac{[\cos(2\pi k_2/L)+\cos(2\pi k_3/L)]^n}{\sqrt{\left(1-\frac{1}{3}[\cos(2\pi k_2/L)+\cos(2\pi k_3/L)]\right)^2-\frac{1}{9}}}  \\
T & = & \frac{1}{L^2}\sum_{(k_2,k_3)\neq(0,0)}^{L-1}\frac{\cos(2\pi k_2/L)\cos(2\pi k_3/L)}{\sqrt{\left(1-\frac{1}{3}[\cos(2\pi k_2/L)+\cos(2\pi k_3/L)]\right)^2-\frac{1}{9}}} .
\end{eqnarray}

We use these propagators to express the matrices $\mathbf{P}$ and $\mathbf{A}$, and obtain the expression of $\Omega$. Finally,
\begin{equation}
\label{general}
\lim_{\rho_0\to 0}\frac{\sigma_x^2}{\rho_0} \underset{t \to \infty}{\sim} \frac{4a_0^2}{L^2}\sqrt{\frac{2}{3\pi}} t^{3/2}.
\end{equation}

\subsubsection{Two-dimensional infinite lattice}
\label{2D_infinite}

In this geometry, the matrices $\mathbf{A}$ and $\mathbf{P}$ are \cite{PRE2002} :
\begin{equation}
\label{AA}
{\rm {\bf A}}=
\begin{pmatrix}
a& \delta q_1 f& \delta q_{-1} f& \delta q_2 f& \delta q_2 f& \\
b& 0    & \delta q_{-1} e& \delta q_2 c& \delta q_2 c& \\
b& \delta q_1 e& 0      & \delta q_2 c& \delta q_2 c& \\
b& \delta q_1 c& \delta q_{-1} c& 0    & \delta q_2 e& \\
b& \delta q_1 c& \delta q_{-1} c& \delta q_2 e& 0,
\end{pmatrix},
\end{equation}
where 
\begin{eqnarray}
\label{aa}
a&\equiv&1-(1-\xi)G(\xi),\;\;\;\;b\equiv\frac{1-\xi}{\xi}(1-G(\xi)),\;\;\;\;e\equiv\frac{4}{3}(2g(\xi)-1),\nonumber\\
c&\equiv&\frac{4}{3}\left(-1+\frac{2}{\xi^2}+2G(\xi)\left(1-\frac{1}{\xi^2}\right)-g(\xi)\right),
\end{eqnarray}
and 
\begin{equation}
\label{PP}
{\rm {\bf P}}=
\begin{pmatrix}
G(\xi)& (G(\xi)-1)/\xi& (G(\xi)-1)/\xi& (G(\xi)-1)/\xi&(G(\xi)-1)/\xi\\
(G(\xi)-1)/\xi& G(\xi)& G(\xi)-2g(\xi)& \tau(\xi)& \tau(\xi)\\
(G(\xi)-1)/\xi&  G(\xi)-2g(\xi)& G(\xi)& \tau(\xi)& \tau(\xi)\\
(G(\xi)-1)/\xi&  \tau(\xi)& \tau(\xi)& G(\xi)& G(\xi)-2g(\xi)\\
(G(\xi)-1)/\xi&  \tau(\xi)& \tau(\xi)&  G(\xi)-2g(\xi)& G(\xi)
\end{pmatrix},
\end{equation}
with
\begin{eqnarray}
G(\xi)&\equiv& P({\bf 0}\;|\;{\bf 0};\xi),\;\;\;g(\xi)\equiv-\frac{1}{2}\left(P({\bf
e_1}\;|\;{\bf -e_1};\xi)-P({\bf 0}\;|\;{\bf
0};\xi)\right),\;\;\; \nonumber\\
\tau(\xi)&\equiv& \left(\frac{2}{\xi^2}-1\right)G(\xi)-\frac{2}{\xi^2}+g(\xi).
\end{eqnarray}
The functions $G$ and $g$ have the following expansions in the $\xi\to 1^-$ limit
\begin{eqnarray}
G(\xi) & \underset{\xi \to 1^-}{=} & \frac{1}{\pi}\ln \frac{1}{1-\xi} + \frac{\ln 8}{\pi} - \frac{1}{2\pi} (1-\xi)\ln(1-\xi) + \mathcal{O}(1-\xi) \\
g(\xi) & \underset{\xi \to 1^-}{=} &  \left( 2-\frac{4}{\pi}\right) + \frac{2}{\pi} (1-\xi)\ln(1-\xi) + \mathcal{O}(1-\xi).
\end{eqnarray}

Using the defintion of $\Omega$ from Eq. (\ref{Omega}), it is possible to show that 
\begin{equation}
\label{general}
\lim_{\rho_0\to 0}\frac{\sigma_x^2}{\rho_0}  \underset{t \to \infty}{\sim} a_0 t \left[\frac{2 a_0}{\pi} \ln t+\frac{2 a_0}{\pi}(\ln 8+\gamma-1) + 2 a_0  \frac{\pi(5-2\pi)}{2\pi-4}  +\coth( f/2) \right], 
\end{equation}
where $\gamma$ is the Euler-Mascheroni constant.


\subsubsection{Three-dimensional infinite lattice}

In this situation, the propagators of an unbiased random walk are particularly simple, since they do not depend on $\xi$ at the leading order. Following \cite{Hughes},
\begin{equation}
\label{propsym3D}
\widehat{P}(\mathbf{r}|\zz ; \xi)=\int_0^\infty \ex{-t} \prod_{j=1}^3 I_{|r_j|} (t\xi/3) \dd t
\end{equation}
where $I_k$ is the modified Bessel function of the first kind of order $k$. It is then easy to express the matrices $\mathbf{P}$ and $\mathbf{A}$. Finally,
\begin{equation}
\label{var_3d}
\lim_{\rho_0\to 0}\frac{\sigma_x^2}{\rho_0} \underset{t \to \infty}{\sim} a_0 t  \left[\coth( f/2)  +2a_0 \left(  P(\mathbf{0}|\mathbf{0} ;1)  + \frac{2(13\alpha-6)}{(2+\alpha)(\alpha-6)} \right)  \right].
\end{equation}
$P(\mathbf{0}|\mathbf{0} ;1)$ is explicitely known (see \cite{Hughes}), and $\alpha$ can be numerically evaluated using Eq. (\ref{propsym3D}). Since $\sigma_x^2$ grows linearly in time, there is no superdiffusion in this geometry.

\section{Giant diffusion regime}

We now consider the other limit, in which we first take $\xi \rightarrow 1^-$ (large time limit) and ultimately take $\rho_0 \rightarrow 0$. The key point is that the vacancy random walk between two successive visits of the lattice site occupied by the TP can be viewed as a biased random walk, with the following jump probabilities ($\widetilde{p}_\mu$ denotes the probability for the vacancy to jump in the $\emuu$ direction) :
\begin{equation}
\label{ptilde}
\widetilde{p}_\mu=\widetilde{p}(\emuu|\zz)\equiv
\begin{cases}
\frac{1/2d+\epsilon}{1+\epsilon}& \text{if $\mu=-1$},\\
\frac{1}{2d(1+\epsilon)}& \text{if $\mu \neq -1$}.
\end{cases}
\end{equation} 
The bias $\epsilon$ originates from the displacement of the TP in the $\epu$ direction, and is the assumed to vanish when $\rho_0$ goes to zero.

The method presented in the previous section can then be applied, replacing the symmetric propagators $\widehat{P}$ by biased propagators corresponding to the jump rules defined in Eq. (\ref{ptilde}), depending on $\epsilon$ and $\xi$, and denoted as $\widehat{\mathcal{P}}(\mathbf{r}|\zz ; \epsilon, \xi)$. In what follows, we present the expressions of the biased propagators in the different geometries, and the resulting expressions of the variance given in the main text.

\subsection{Two-dimensional strip}

The general expression of the propagator of a biased random walk on a strip-like lattice is
\begin{equation}
\widehat{\mathcal{P}}(\mathbf{r}|\zz;\xi,\epsilon) =  \frac{1}{L}\sum_{k=0}^{L-1} \frac{1}{2\pi} \ex{-2i\pi k n_2/L} \underbrace{\int_{-\pi}^\pi \dd q \frac{\ex{-i n_1 q} }{1-\xi (\widetilde{p}_1\ex{iq}+\widetilde{p}_{-1}\ex{-iq}+2\widetilde{p}_2\cos(2\pi k/L))}}_{\equiv f_2(n_1,k)},
\end{equation}
where $\mathbf{r}=n_1 \mathbf{e_1} + n_2 \mathbf{e_2} $.  One can derive an explicit expression of $\widehat{\mathcal{P}}$ with the change of variable $u=\ex{-iq}$ and using the residue theorem :
\begin{equation}
f_2(n_1,k)=
\begin{cases}
- \frac{2\pi}{\xi \widetilde{p}_{-1}} \frac{U_2^{n_1}}{U_2-U_1} & \text{if $n_1 \geq 0$},\\
- \frac{2\pi}{\xi \widetilde{p}_{-1}} \frac{1}{U_1^{|n_1|}(U_2-U_1)} & \text{if $n_1 < 0$}.
\end{cases}
\end{equation} 
with 
\begin{equation}
\label{ }
U_{1,2} = \frac{1}{2} \left\{ \frac{1}{\xi \widetilde{p}_{-1}} \left( 1-2\xi \widetilde{p}_{2}\cos \frac{2\pi k}{L}\right) \pm \sqrt{\frac{1}{(\xi \widetilde{p}_{-1})^2}\left( 1-2\xi \widetilde{p}_{2}\cos \frac{2\pi k}{L}\right)^2  -4 \frac{\widetilde{p}_1}{\widetilde{p}_{-1}} }  \right\}.
\end{equation}

It is then possible to calculate the matrix $\mathbf{P^+}$, the functions $\widehat{F}^*({\bf 0}|{\bf e}_{\boldsymbol \mu}|{\bf e}_{\boldsymbol\nu};\xi)$ and the single vacancy propagator $\widehat{\widetilde{P}}^{(1)}({\bf k};\xi)$. Finally, taking first the limit $\xi \rightarrow 1^-$, we get $\Omega$ as a function of $\epsilon$, and find in the leading order in $\rho_0$ :
\begin{equation}
\label{variance_strip_diff}
 \lim_{t\to\infty}  \frac{\sigma_x^2}{t}  \underset{\rho_0 \to 0}{\sim}  \frac{2 a_0'}{L}
\end{equation}
where $a'_0$ only depends on quantities of the form $\nabla_{\mu} \widehat{\mathcal{P}}(\mathbf{r} | \zz, \epsilon=0)$ ($\mu\in \{ \pm 1, \pm 2\}$), the operator $\nabla_\mu$ being defined for any space-dependent function $\varphi$ by
\begin{equation}
\label{ }
\nabla_\mu \varphi (\boldsymbol{\lambda}) = \varphi (\boldsymbol{\lambda}+\emuu)-\varphi (\boldsymbol{\lambda}).
\end{equation}
The quantity $B$ used in the main text is then $B=2a'_0/L$.

Using the explicit derivation of $\widehat{\mathcal{P}}$, we can relate such `gradients' with the symmetric propagators $\widehat{P}$ defined in section \ref{superdiff_strip} :
\begin{equation}
\lim_{\epsilon \to 0} \nabla_\mu \widehat{\mathcal{P}}(\mathbf{r}|\zz;\epsilon) -  \lim_{\xi \to 1} \nabla_\mu \widehat{P}(\mathbf{r}|\zz;\xi) = \left\{
\begin{array}{l l}
-2/L & \mathrm{if~} \mu=1 \\
2/L & \mathrm{if~} \mu=-1 \\
0 & \mathrm{otherwise}
\end{array}
\right.
\end{equation}

Finally, using the symmetry properties of $\widehat{P}$, we find
\begin{equation}
\label{ }
a'_0 = \frac{p_1-p_{-1}}{1+4 [x_+ (p_1+p_{-1}) +  x_- (p_1-p_{-1}) ]}
\end{equation}
where
\begin{eqnarray}
x_+ & = & \frac{\delta_{1,1}-\delta_{-1,1}}{\delta_{-1,1}-\delta_{1,1}-4}  \label{xp}  \\
x_- & = & \frac{1}{L}\frac{16(\delta_{2,2}+\delta_{-2,2}-2\delta_{2,1}-4)}{(\delta_{-1,1}-\delta_{1,1}-4)[(\delta_{2,2}+\delta_{-2,2})(\delta_{-1,1}+\delta_{1,1})-4(\delta_{-1,1}+\delta_{1,1}+\delta_{2,2}+\delta_{-2,2}+\delta_{2,1}\delta_{1,2}-4)]}    \label{xm}
\end{eqnarray}
and
\begin{equation}
\label{ }
\delta_{\mu,\nu} = \lim_{\xi \to 1} \nabla_\mu \widehat{P}(\enu | \zz ; \xi).
\end{equation}

Then quantities $\delta_{\mu, \nu}$ are calculated using Eqs. (\ref{g1}) to (\ref{g6}). Finally, Eq. (\ref{variance_strip_diff}) is fully explicit.

\subsection{Three-dimensional capillary}

The general expression of the propagator of a biased random walk on a strip-like lattice is
\begin{equation}
\widehat{\mathcal{P}}(\mathbf{r}|\zz;\xi, \epsilon)  =  \frac{1}{L^2}\sum_{k_2=0}^{L-1}\sum_{k_3=0}^{L-1} \frac{1}{2\pi} \ex{-2i\pi (k_2 n_2+k_3 n_3)/L} \underbrace{\int_{-\pi}^\pi \dd q \frac{\ex{-i n_1 q} }{1-\xi (\widetilde{p}_1\ex{iq}+\widetilde{p}_{-1}\ex{-iq}+2\widetilde{p}_2\cos(2\pi k_2/L))+2\widetilde{p}_3\cos(2\pi k_3/L))}}_{\equiv f_3(n_1,k)}
\end{equation}

where $\mathbf{r}=n_1\mathbf{e_1}+n_2\mathbf{e_2}+n_3\mathbf{e_3}$. With the change of variable $u=\ex{-iq}$ and using the residue theorem, one shows that
\begin{equation}
f_3(n_1,k)=
\begin{cases}
- \frac{2\pi}{\xi \widetilde{p}_{-1}} \frac{V_2^{n_1}}{V_2-V_1} & \text{if $n_1 \geq 0$},\\
- \frac{2\pi}{\xi \widetilde{p}_{-1}} \frac{1}{V_1^{|n_1|}(V_2-V_1)} & \text{if $n_1 < 0$}.
\end{cases}
\end{equation} 
with 
\begin{equation}
\label{ }
V_{1,2} = \frac{1}{2} \left\{ \frac{1}{\xi \widetilde{p}_{-1}} \left[ 1-2\xi \left(\widetilde{p}_{2}\cos \frac{2\pi k_2}{L}+\widetilde{p}_{3}\cos \frac{2\pi k_3}{L}\right)\right] \pm \sqrt{\frac{1}{(\xi \widetilde{p}_{-1})^2}\left[ 1-2\xi \left(\widetilde{p}_{2}\cos \frac{2\pi k_2}{L}+\widetilde{p}_{3}\cos \frac{2\pi k_3}{L}\right)\right]^2  -4 \frac{\widetilde{p}_1}{\widetilde{p}_{-1}} }  \right\}.
\end{equation}

It is then possible to calculate the matrix $\mathbf{P^+}$, the functions $\widehat{F}^*({\bf 0}|{\bf e}_{\boldsymbol \mu}|{\bf e}_{\boldsymbol\nu};\xi)$ and the single vacancy propagator $\widehat{\widetilde{P}}^{(1)}({\bf k};\xi)$. Finally, taking first the limit $\xi \rightarrow 1^-$, we get $\Omega$ as a function of $\epsilon$, and find in the leading order in $\rho_0$ :
\begin{equation}
\label{variance_cap_diff}
 \lim_{t\to\infty}  \frac{\sigma_x^2}{t}  \underset{\rho_0 \to 0}{\sim}  \frac{\sqrt{6} a_0'}{L^2}
\end{equation}
where $a'_0$ only depends on quantities of the form $\nabla_{\mu} \widehat{\mathcal{P}}(\mathbf{r} | \zz, \epsilon=0)$ ($\mu\in \{ \pm 1, \pm 2, \pm 3\}$). The quantity $B$ used in the main text is then $B=\sqrt{6} a'_0/L^2$.

Using the explicit derivation of $\widehat{\mathcal{P}}$, we can express such `gradients' in terms  the symmetric propagators $\widehat{P}$ defined in the previous section :
\begin{equation}
\lim_{\epsilon \to 0} \nabla_\mu \widehat{\mathcal{P}}(\mathbf{r}|\zz;\epsilon) -  \lim_{\xi \to 1} \nabla_\mu \widehat{P}(\mathbf{r}|\zz;\xi) = \left\{
\begin{array}{l l}
-3/L^2 & \mathrm{if~} \mu=1 \\
3/L^2 & \mathrm{if~} \mu=-1 \\
0 & \mathrm{otherwise}
\end{array}
\right.
\end{equation}

Finally, using the symmetry properties of $\widehat{P}$, we find
\begin{equation}
\label{ }
a'_0 = \frac{p_1-p_{-1}}{1+6 [x_+ (p_1+p_{-1}) +  x_- (p_1-p_{-1}) ]}
\end{equation}
where
\begin{eqnarray}
x_+ & = & \frac{\delta_{1,1}-\delta_{-1,1}}{\delta_{-1,1}-\delta_{1,1}-6}  \label{xp}  \\
x_- & = & \frac{1}{L^2}\frac{36(\delta_{2,2}+\delta_{-2,2}+2\delta_{3,2}-4\delta_{2,1}-6)}{(\delta_{-1,1}-\delta_{1,1}-6)[(\delta_{2,2}+\delta_{-2,2})(\delta_{-1,1}+\delta_{1,1}+2\delta_{3,2})-6(\delta_{-1,1}+\delta_{1,1}+\delta_{2,2}+\delta_{-2,2}+2\delta_{3,2}+4\delta_{2,1}\delta_{1,2}/3-6)]}    \label{xm}\nonumber\\
\end{eqnarray}
and
\begin{equation}
\label{ }
\delta_{\mu,\nu} = \lim_{\xi \to 1} \nabla_\mu \widehat{P}(\enu | \zz ; \xi).
\end{equation}

Then quantities $\delta_{\mu, \nu}$ are calculated using Eqs. (\ref{h1}) to (\ref{h7}). Finally, Eq. (\ref{variance_cap_diff}) is fully explicit.

\subsection{Two-dimensional infinite lattice}
\label{2Ddiff}

The  propagator can be written as :
\begin{equation}
\widehat{\mathcal{P}}(\mathbf{r}|\zz ; \epsilon, \xi=1)=\frac{1}{(2\pi)^2}\int_{-\pi}^\pi  \dd k \int_{-\pi}^\pi \dd q \frac{\ex{-iqn_1-ikn_2}}{1-\left(\widetilde{p}_1\ex{iq}   +\widetilde{p}_{-1}\ex{-iq} + 2\widetilde{p}_2 \cos k   \right)}
\end{equation}
We then remark that
\begin{eqnarray}
\label{ }
\lim_{\epsilon \to 0} \left[  \widehat{\mathcal{P}}(\mathbf{r}|\zz ; \epsilon, \xi=1) - \widehat{\mathcal{P}}(\zz|\zz ; \epsilon, \xi=1) \right] &=& \lim_{\xi \to 1} \left[  \widehat{\mathcal{P}}(\mathbf{r}|\zz ; \epsilon=0, \xi) - \widehat{\mathcal{P}}(\zz|\zz ; \epsilon=0, \xi) \right]\\
&=&  \lim_{\xi \to 1} \left[  \widehat{{P}}(\mathbf{r}|\zz ; \xi) - \widehat{{P}}(\zz|\zz ; \xi) \right].
\end{eqnarray}
Recalling that the propagators $\widehat{P}$ can be easily calculated, we write
\begin{equation}
\label{2Dbiasedprop}
\widehat{\mathcal{P}}(\mathbf{r}|\zz ; \epsilon, \xi=1) \underset{\epsilon \to 0}{\sim} \widehat{\mathcal{P}}(\zz|\zz ; \epsilon, \xi=1)+\lim_{\xi \to 1} \left[  \widehat{{P}}(\mathbf{r}|\zz ; \xi) - \widehat{{P}}(\zz|\zz ; \xi) \right] .
\end{equation}

Next, $ \widehat{\mathcal{P}}(\zz|\zz ; \epsilon, \xi=1)$ can be calculated as follows :
\begin{eqnarray}
\widehat{\mathcal{P}}(\zz|\zz ; \epsilon, \xi=1) & = & \frac{1}{(2\pi)^2}\int_{-\pi}^\pi  \dd k \int_{-\pi}^\pi \dd q \frac{1}{1-\left(\widetilde{p}_1\ex{iq}   +\widetilde{p}_{-1}\ex{-iq} + 2\widetilde{p}_2 \cos k   \right)} \\
 & = & \frac{1}{\pi} \int_0^\pi \dd k \frac{1}{\sqrt{(1-2\widetilde{p}_2\cos k)^2-4\widetilde{p}_1\widetilde{p}_{-1}}}\mathrm{~~~(residue~theorem)}\\
 & = & \frac{1}{2\pi\widetilde{p}_2} \frac{2}{\sqrt{a+1}\sqrt{b-1}} K\left( \sqrt{\frac{2(b-a)}{(a+1)(b-1)}}\right) 
\end{eqnarray}
where $a=(1-2\sqrt{\widetilde{p}_1\widetilde{p}_{-1}})/2\widetilde{p}_2$, $b=(1+2\sqrt{\widetilde{p}_1\widetilde{p}_{-1}})/2\widetilde{p}_2$ and $K$ is the complete elliptic integral of the first kind. Using the known expansion of $K$ close to $1$, we find 
\begin{equation}
\label{ }
\widehat{\mathcal{P}}(\zz|\zz ; \epsilon, \xi=1) \underset{\epsilon \to 0}{\sim} \frac{2}{\pi} \ln \frac{1}{\epsilon} + \frac{1}{\pi} \ln 8 + \mathcal{O}(\epsilon^2 \ln \epsilon ) .
\end{equation}

Using this expansion and Eq. (\ref{2Dbiasedprop}), we can calculate all intermediate quantities and finally get the following expression of the variance :
\begin{equation}
\label{general}
\lim_{t\to\infty}\frac{\sigma_x^2}{t} \underset{\rho_0 \to 0}{\sim}  \rho_0 a_0 \left[ \coth( f/2)+ 2 a_0  \left( \frac{2}{\pi} \ln \frac{1}{\rho_0 a_0}+\frac{1}{\pi} \ln 8+   \frac{\pi(5-2\pi)}{2\pi-4}  \right) \right].
\end{equation}

\subsection{Three-dimensional infinite lattice}

For 3D lattices, the leading order term of the propagators is the same in the two limits ($\xi \to 1$, $\epsilon \to 0$) and ($\epsilon \to 0$, $\xi \to 1$), so that the variance is still given by Eq. (\ref{var_3d}).

\section{Scaling regime and cross-over}

Finally, we describe the intermediate time regime ($t \simeq t_\times$, see main text) by considering the scaling limit where $1-\xi$ and $\epsilon$ simultaneously tend to zero with the scaling law $1-\xi \sim \epsilon^2$.

\subsection{Two-dimensional strip}

Following the same method, we  get the following expression of the variance in the scaling limit :
\begin{equation}
\label{ }
\widehat{\sigma}_x^2(\epsilon,\xi) = \frac{\rho_0}{(1-\xi)^2}2 a_0(\epsilon,\xi)^2 \widehat{\mathcal{P}}(\zz|\zz ; \epsilon, \xi) +\mathcal{O}(1-\xi,\epsilon)
\end{equation}
where 
\begin{equation}
\label{ }
\widehat{\mathcal{P}}(\zz|\zz ; \epsilon, \xi) = \frac{1}{L\sqrt{1-\xi+\epsilon^2}} +\mathcal{O}(1-\xi,\epsilon)
\end{equation}
and $a_0(\epsilon,\xi)$ has the form
\begin{equation}
\label{ }
a_0(\epsilon,\xi) = \frac{p_1-p_{-1}}{1+4 \left[x_+ (p_1+p_{-1}) +  x_-  \frac{\epsilon}{\sqrt{1-\xi+\epsilon^2}} (p_1-p_{-1}) \right]} +\mathcal{O}(1-\xi,\epsilon)
\end{equation}
where $x_+$ and $x_-$ were defined by Eqs. (\ref{xp}), (\ref{xm}).

These expressions were obtained by using the following relations between the expansions of $\widehat{\mathcal{P}}$ and the expansions of $\widehat{P}$ in the joint limit :
\begin{equation}
\nabla_\mu \widehat{\mathcal{P}}(\mathbf{r}|\zz;\epsilon, \xi) -  \lim_{\xi \to 1} \nabla_\mu \widehat{P}(\mathbf{r}|\zz;\xi) = \left\{
\begin{array}{l l}
-\frac{2}{L} \frac{\epsilon}{\sqrt{1-\xi+\epsilon^2}} +\mathcal{O}(1-\xi,\epsilon) & \mathrm{if~} \mu=1 \\
\frac{2}{L} \frac{\epsilon}{\sqrt{1-\xi+\epsilon^2}} +\mathcal{O}(1-\xi,\epsilon) & \mathrm{if~} \mu=-1 \\
\mathcal{O}(1-\xi,\epsilon) & \mathrm{otherwise.}
\end{array}
\right.
\end{equation}

Finally, we get the following expression for the variance
\begin{equation}
\label{ }
\widehat{\sigma}_x^2(\epsilon,\xi) \sim \frac{2 \rho_0}{L\epsilon} \left( \frac{p_1-p_{-1}}{1+ 4 x_+(p_1+p_{-1})}  \right)^2  {\frac{1}{(1-\xi)^2} \frac{\sqrt{1+\frac{1-\xi}{\epsilon^2}}}{\left(  \frac{4 x_- (p_1-p_{-1})}{1+4 x_+(p_1+p_{-1})}    +\sqrt{1+\frac{1-\xi}{\epsilon^2}}\right)^2}}.
\end{equation}
In order to obtain an expression of the variance as a function of the initial parameters only, we replace $\epsilon$ by $\rho_0 a'_0$ (this equivalence is given by the explicit calculation of the TP velocity, not presented here). We finally obtain the scaling law
\begin{equation}
\label{ }
\sigma_x^2(t) \sim t g( \rho_0^2 t)
\end{equation}
with
\begin{equation}
\label{ }
\begin{split}
g(x)&=\frac{1}{L {a'_0}^3(\zeta^2-1)^3 x} \left( \frac{p_1-p_{-1}}{1+ 4 x_+(p_1+p_{-1})}  \right)^2 \Big\{  4\zeta({a'_0}^2 x(1-{a'_0}^3 x)-\zeta(1+\zeta))   \\
&+ \mathrm{erf}({a'_0}\sqrt{x})(2{a'_0}^2 x(\zeta^4-1)+\zeta^4+6\zeta^2+1)+2\ex{-{a'_0}^2 x}{a'_0}\sqrt{\frac{x}{\pi}}(\zeta-1)(\zeta+1)(3\zeta^2+1)\\
&+\ex{{a'_0}^2 (\zeta^2-1)x}\mathrm{erfc}({a'_0}\zeta\sqrt{x})4\zeta(4\zeta^2(1+{a'_0}^2 x)+1-\zeta^4 {a'_0}^2 x) \Big\}
\end{split}
\end{equation}
and
\begin{eqnarray}
\label{ }
\zeta&=&\frac{4x_- (p_1-p_{-1})}{1+4 x_+(p_1+p_{-1})}\\
\mathrm{erf}(X)&=&\frac{2}{\sqrt{\pi}}\int_0^X\ex{-t^2}\dd t \mathrm{~~~~~~~(error~function)}\\
\mathrm{erfc}(X)&=&1- \mathrm{erf}(X)  
\end{eqnarray}

\subsection{Two-dimensional infinite lattice}

Eq. (\ref{2Dbiasedprop}), which was derived in the ($\xi \to 1$, $\epsilon \to 0$) limit, is still valid in the joint limit, so that we only need the expansion of $\widehat{\mathcal{P}}(\mathbf{r}|\zz ; \epsilon, \xi)$ in the joint limit. Using the definition
\begin{equation}
\widehat{\mathcal{P}}(\mathbf{r}|\zz ; \epsilon, \xi)=\frac{1}{(2\pi)^2}\int_{-\pi}^\pi  \dd k \int_{-\pi}^\pi \dd q \frac{\ex{-iqn_1-ikn_2}}{1-\xi \left(\widetilde{p}_1\ex{iq}   +\widetilde{p}_{-1}\ex{-iq} + 2\widetilde{p}_2 \cos k   \right)}
\end{equation}
and the procedure used in the $\xi\to 1$ limit, we find the following expansion in the joint limit
\begin{equation}
\widehat{\mathcal{P}}(\zz|\zz ; \epsilon, \xi)= \frac{1}{\pi} \ln \frac{1}{1-\xi+\epsilon^2} + \frac{1}{\pi} \ln 8 + \mathcal{O}(1-\xi, \epsilon).
\end{equation}
We then obtain the following expression for the variance in the joint limit
\begin{equation}
\label{ }
\widehat{\sigma}_x^2(\epsilon,\xi) = \frac{\rho_0 a_0}{(1-\xi)^2}\left[\coth( f/2) +2a_0\left( \widehat{\mathcal{P}}(\zz|\zz ; \epsilon, \xi) + \frac{\pi(5-2\pi)}{2\pi-4} \right) \right]+\mathcal{O}(1-\xi,\epsilon).
\end{equation}
In order to obtain an expression of the variance as a function of the initial parameters only, we replace $\epsilon$ by $\rho_0 a'_0$(this equivalence is given by the explicit calculation of the TP velocity, not presented here). We finally invert the expression of the variance in order to obtain the following time-dependent result :
\begin{equation}
\label{ }
\sigma_x^2(t) \sim \rho_0 a_0 t \left[ \frac{2 a_0}{\pi} \ln \frac{t \rho_0^2 a_0^2}{1+t \rho_0^2 a_0^2} +   \coth( f/2) + 2a_0  \frac{\pi(5-2\pi)}{2\pi-4} + \frac{2a_0}{\pi}\left( \ln 8 +\gamma-1 + \ln \frac{1}{\rho_0^2 a_0^2}  \right) \right].
\end{equation}

\section{Continuous-time random walk (CTRW) description}

\subsection{Introduction}

We now focus on a simplified continuous-time description where the random walk of the TP is completely directed (i.e. $f\rightarrow \infty$, $p_1=1$, $p_{\nu \neq 1}=0$) so that the TP may only jump in the direction of the unitary vector $\mathbf{e_1}$. As previously, $\mathbf{X}$ is the random variable representing the position of the TP at time $t$, and $X_t$ its projection of the $x$-axis. We also assume that when a vacancy is at position $\mathbf{X}+\mathbf{e_1}$ at time $t$, then the TP instantly jumps to position $\mathbf{X}+\mathbf{e_1}$, exchanging its position with the vacancy. We denote as $N_t$ the number of times that the site located at the right of the TP was visited by a vacancy between times $0$ and $t$. Under the previous assumption, we conclude that the random variables $X_t$ and $N_t$ have the same properties. We can then study the TP position by describing $N_t$, assuming that the vacancies perform independent random walks on the lattice.
This approach also relies on the mean-field assumption that after each interaction of the TP with a vacancy, all other vacancies are uniformly distributed.

\subsection{General expressions of the variance and leading behaviors}

We assume that $M$ independent vacancies are homogeneously distributed on a $N$-site lattice, and that the TP is initially at site $\mathbf{e_{-1}}$. Denoting as $F(t|\mathbf{r_0})$ the probability that a vacancy starting its random walk at site $\mathbf{r_0}$ visits the origin for the first time at time $t$, we show that
\begin{equation}
\label{ }
S_1(t)=\left[1-\frac{1}{M}\sum_{\mathbf{y_0}\neq \mathbf{0}}\sum_{n=0}^tF^*_n(\zz|\mathbf{e_1}|\mathbf{y_0}) \right]^N
\end{equation}
where $S_1(t)$ is the probability that the origin was not visited yet at time $t$ (known as survival probability).
In the limit $M\rightarrow\infty$, $N\rightarrow\infty$ with fixed $\rho_0=M/N$,
\begin{equation}
\label{ }
S_1(t)=\exp\left[-\rho_0 \sum_{\mathbf{y_0}\neq \mathbf{0}}\sum_{n=0}^tF^*_n(\zz|\mathbf{e_1}|\mathbf{y_0}) \right]
\end{equation}
In what follows, we define
\begin{equation}
\label{ }
G(t)\equiv  \sum_{\mathbf{y_0}\neq \mathbf{0}}\sum_{n=0}^tF^*_n(\zz|\mathbf{e_1}|\mathbf{y_0}).
\end{equation}

The calculation relies on the following mean-field-like approximation. After the first visit of the TP, the system is set in a new initial condition : a vacancy is at site $\mathbf{0}$, whereas the rest of the lattice is homogeneously occupied by vacancies with density $\rho_0$. The survival probality then becomes
\begin{equation}
\label{ }
S(t) = T(t)S_1(t)
\end{equation}
where $T(t)$ is the survival probability associated to a TP at site $\mathbf{e_1}$ and a vacancy at site $\mathbf{0}$. Let $\psi_1(t)$ be the waiting time distribution before the first jump occurs, and $\psi(t)$ the waiting time distributions between two consecutive jumps occuring after the first one. Note that these quantities are related to the survival probabilities through the relations
\begin{eqnarray}
\psi_1(t) &=& - \frac{\dd S_1}{\dd t} \\
\psi (t) &=& - \frac{\dd S}{\dd t}
\end{eqnarray}

Using the notations of the previous section, we find that
\begin{equation}
\label{ }
T(t) = 1-\sum_{n=0}^tF^*_n(\zz|\mathbf{e_1} | \mathbf{e_{-1}})
\end{equation}
which is equivalent to
\begin{equation}
\label{ }
\widehat{T}(\xi)=\frac{1}{1-\xi}[1-F^*(\zz|\mathbf{e_1} | \mathbf{e_{-1}};\xi)].
\end{equation}

Using these definitions, and denoting with $\widehat{\cdot}$ the Laplace transform, we show that
\begin{eqnarray}
\widehat{\moy{N_t}}(\xi) & = & \frac{1}{(1-\xi)^2} \frac{1-(1-\xi)\widehat{S_1}(\xi)}{\widehat{S}(\xi)} \\ 
\widehat{\moy{N^2_t}}(\xi) &=&  \frac{1}{(1-\xi)^3} \frac{[1-(1-\xi)\widehat{S_1}(\xi)][2-(1-\xi)\widehat{S}(\xi)]}{(\widehat{S}(\xi))^2}  \label{moy2s}
\end{eqnarray}

In the particular case of a strip-like geometry, and using the previous results, we show 
\begin{equation}
\label{ }
\sum_{n=0}^t \widehat{F}^*(\zz|\mathbf{e_1}|\mathbf{y_0} ; \xi) \underset{\xi \to 1^-}{\sim} \frac{L}{\sqrt{1-\xi}},
\end{equation}
so that
\begin{equation}
\label{ }
\widehat{G}(\xi) \underset{\xi \to 1^-}{\sim} \frac{L}{(1-\xi)^{3/2}}\mathrm{~~~ and ~~~} G(t) \underset{t\to + \infty}{\sim}\frac{2L}{\sqrt{\pi}}\sqrt{t}.
\end{equation}
Similarly,
\begin{equation}
\label{ }
\widehat{F}^*(\zz|\mathbf{e_1} | \mathbf{e_{-1}};\xi)=1-\frac{L}{a_0}\sqrt{1-\xi}+\mathcal{O}(1-\xi),
\end{equation}
so that
\begin{equation}
\label{ }
\widehat{T}(\xi) \underset{\xi \to 1^-}{\sim} \frac{L}{a_0}\frac{1}{\sqrt{1-\xi}}\mathrm{~~~ and ~~~} T(t) \underset{t\to + \infty}{\sim} \frac{L}{a_0}\frac{1}{\sqrt{\pi t}}.
\end{equation}

\subsubsection{Long-time behavior of the low vacancy density limit}

In the $\rho_0 \to 0$ limit :
\begin{eqnarray}
\widehat{S}_1(\xi)&=&\sum_{t=0}^{+\infty}\xi^t{\ex{-\rho_0 G(t)}}  =  \sum_{t=0}^{+\infty}\xi^t [1-\rho_0 G(t)+\mathcal{O}(\rho_0^2)] = \frac{1}{1-\xi}-\rho_0\widehat{G}(\xi)+\mathcal{O}(\rho_0^2) \\
\widehat{S}(\xi)&=&\sum_{t=0}^{+\infty}\xi^t  T(t) {\ex{-\rho_0 G(t)}} = \widehat{T}(\xi)+\mathcal{O}(\rho_0) \label{S_superdiff}
\end{eqnarray}

Using these expansions, we get
\begin{equation}
\lim_{\rho_0 \to 0} \frac{\mathrm{Var}(N(t)) }{\rho_0} \underset{t \to \infty}{\sim}  \frac{2a_0^2}{L}\frac{4}{3\sqrt{\pi}}t^{3/2}.
\end{equation}
which is equivalent to Eq. (\ref{varsuperdiffstrip}). Note that this superdiffusive behavior is due to the asymptotic behavior of $T$ (Eq. (\ref{S_superdiff})), which describes the survival probability of an immobile target purchased by a \emph{single} random walker starting near the target.

\subsubsection{Low vacancy density behavior of the long time limit}

In the $\xi \to 1^-$ limit :

\begin{eqnarray}
\widehat{S}_1(\xi)&=&\sum_{t=0}^{+\infty}\xi^t S_1(t)  = \underbrace{ \sum_{t=0}^{+\infty}S_1(t)}_{\equiv a_1(\rho_0)} - (1-\xi)\underbrace{ \sum_{t=0}^{+\infty}tS_1(t)}_{\equiv b(\rho_0)}+\mathcal{O}\left[(1-\xi)^2\right] \\
\widehat{T}(\xi)&=&\underbrace{ \sum_{t=0}^{+\infty}S(t)}_{\equiv a(\rho_0)} - (1-\xi)\underbrace{ \sum_{t=0}^{+\infty}tS_1(t)}_{\equiv b(\rho_0)}+\mathcal{O}\left[(1-\xi)^2\right]
      \end{eqnarray}

It is found that
\begin{equation}
\label{ }
\lim_{t\to \infty}  \mathrm{Var}(N_t) \underset{\rho_0 \to 0}{\sim} \frac{4b/a-2a_1-a}{a^2}t
\end{equation}
where
\begin{eqnarray}
a_1(\rho_0) & \underset{\rho_0 \to 0}{\sim} & \frac{\pi}{2 L^2 \rho_0^2} \\
a(\rho_0) & \underset{\rho_0 \to 0}{\sim} & \frac{1}{\rho_0 a_0} \\
b(\rho_0) & \underset{\rho_0 \to 0}{\sim} & \frac{\pi}{2 a_0 L^2 \rho_0^3}
\end{eqnarray}
so that
\begin{equation}
\label{ }
\lim_{t\to \infty}  \mathrm{Var}(N_t) \underset{\rho_0 \to 0}{\sim} \frac{\pi a_0^2}{L^2}t.
\end{equation}
Note that the ultimate diffusive behavior is recovered when, after a cross-over time, other vacancies eventually interact with the TP. Although this result is qualitatively in agreement with the exact approach, it does not predict the right dependence with the width of the strip, which is supposed to vary as $1/L$.

\subsection{Conclusions and remarks}

\begin{figure}
	\includegraphics[width = 12cm]{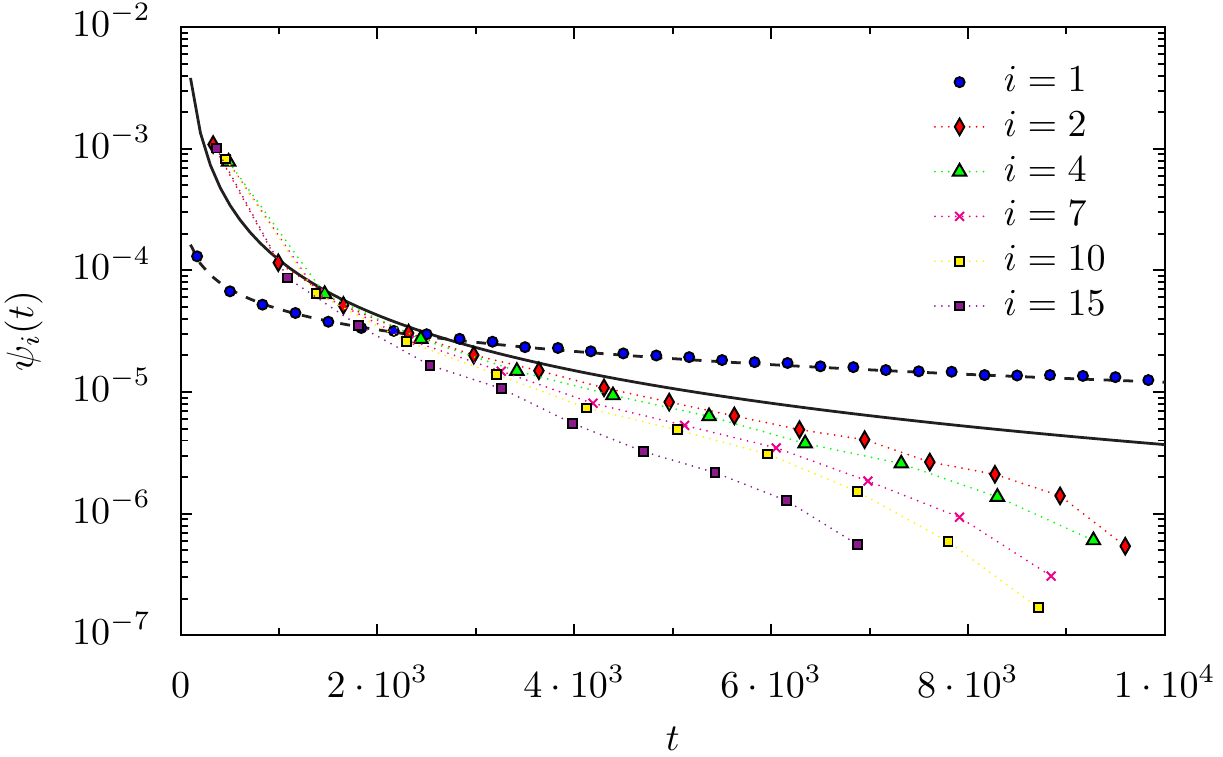}
	\caption{Waiting-times extracted from simulations on a 2D strip ($L=3$, $\rho_0 = 10^{-3}$). Symbols : waiting-time distribution between the $i$-th jump of the TP and the previous jump. Black lines : theoretical expressions of $\psi_1(t)$ (dashed) and $\psi(t)$ (solid).}
	\label{CTRW_simu}
\end{figure}

In the particular case of a strip-like geometry, this CTRW approach allows us to retrieve the two leading behaviors (superdiffusive and diffusive) with $t\to\infty$ and $\rho_0 \to 0$ obtained with the exact lattice gas model. However, although the superdiffusive is in quantitative agreement with the exact result, the prediction of the diffusive regime is not quantitatively correct, and cannot be compared to the results obtained by simulation or exact calculation. Indeed, the mean-field approximation, which is equivalent to considering that the survival probability remains identically distributed once the first jump of the TP occurred is wrong according to our simulation results. Fig. \ref{CTRW_simu} shows that the distribution of the waiting time before the first jump is correctly described by its theoretical expression. However, the next waiting times $\psi_{i>1}(t)$ are not identically distributed. Consequently, they do not correspond to the calculated expression of $\psi(t)$, which is deduced from the mean-field approximation. 

\section{Simulation methods}

\subsection{On-Lattice gas simulations}

To simulate a lattice gas system, we performed Monte-Carlo simulations of $N$ vacancies on an $M$-site lattice ($\rho_0 =M/N$) of size $L_x\times L_y$ (representing 2D) or $L_x\times L_y\times L_z$ (representing 3D), including periodic boundary conditions. To begin with, $N$ vacancies were randomly initiated on the lattice. The system is considered as infinite if a vacancy does not have the time to visit on average all the lattice sites. Per simulation run, the number of distinct sites visited during time $t$ by a vacancy is of the order of $t$. With this knowledge, we impose the number of lattice sites to be at minimum, equal to the the number simulation sites. This guarantees that the vacancies can not visit every lattice site, therefore avoiding finite-size effects.
At each time step, each of the $N$ vacancies exchanges its position with one of the neighboring particles according to the criteria defined in section \ref{Model}. Consequently the TP may move as a result of the vacancies displacements, allowing a  calculation of the relative mean position of the TP, and the corresponding fluctuations.

\subsection{Off-lattice}

In order to investigate numerically the `off-lattice' motion of a colloidal particle
dragged through a crowded environment located in a thin, narrow
channel (i.e., a strip), we performed Molecular Dynamics simulations of two separate systems:
a Lennard-Jones (LJ) system, representing fluid-like systems, and a pure Hard-Disk (HD), representing a granular fluid (GF).

The LJ simulations are composed purely of monodisperse `Lennard--Jones'
particles in 2D. Each particle interacts through a
steep, purely repulsive Lennard --Jones potential described through the equation $U(r)=\epsilon (r/\sigma)^{-48}$, where
$r$ is the inter-particle separation between two interacting
particles, $\sigma$ is the diameter of the LJ particles, $\epsilon$ is
the energy scale, and the time step of the simulation, defined in
units of $\tau=\sqrt{m\sigma^2/\epsilon}$.  The required `strip'
geometry is created through fixing the aspect ratio, $L_y/L_x$ of the
simulation box at very small values, with periodic boundaries along
both the directions.  Specifically, in the LJ simulations, we fix the
aspect ratio at 3 different values $L_y/L_x=0.0017$, $0.0037$ and
$0.0051$, in order to check the dependence of the strip width $L_y$ on
the output of the simulations. Depending on the aspect ratio, the total number of particles within the
simulation varied from 2048 to 6144. In addition, the force acting on
the particles was changed according to $f_0\epsilon/\sigma=1,2,3,4$ and $10$ in order to check
the dependence of the applied force on the fluctuations of the location relating to the TP. 
The system was allowed to reach equilibrium in the NVT ensemble at a
temperature of $1.0 k_{\rm B}T/\epsilon$, using a N\'ose--Hoover
thermostat coupled to the system \cite{Melchionna1993}. After an
initial equilibration time of $2\times10^3 \tau$, the thermostat was
switched off and a further equilibration time of $2\times10^3 \tau$ in the
hydrodynamics preserving NVE ensemble was performed; therefore, hydrodynamics are included via conserving local momentum. Subsequently, a constant force was added to one of the LJ particles, thus creating the TP. 



We also perform pure HD simulations representing a GF in the spirit of \cite{Zippelius}: They each consist of 3000 granular particles of mass $m=1$ and radius $R=1$, placed like LJ simulations, within a 2D strip of varying aspect ratios $L_y/L_x=0.002$, $0.003$ and $0.004$. The particles collisions are dissipative.
In addition, a frictional force defined through $F_{fr} = - \gamma
m\textbf{v}$ acts on all particles. In order to balance the
dissipation, the system is driven stochastically with random force
$F_r$, such that
\begin{equation}
  \left\langle F_{ri}(t)F_{rj}(t^{\,\prime})\right\rangle=\delta_{ij}\delta(t-t^{\,\prime})m^2\xi_0^2.
\label{Fstoch}
\end{equation}

where $\gamma= 1, \xi_0 =150$.
Likewise to the LJ simulations, periodic boundary conditions are used. To maintain
thermal equilibrium, an initiation run lasting $1\times 10^6$ cycles is run. 
Following equilibration, the TP is created in a similar manner as already mentioned, 
through the addition of the constant force $f_0$ at every
time interval to only one of the particles. The chosen force values $f_0$ acting on the particle in the GF were $10, 15 $ and $20$. 

Per simulation run, the trajectory of the TP was recorded at equal time intervals, in order
calculate the variance of distribution in the displacement of the TP
in the $x$ direction, $\sigma _x ^2$. The variance, defined as $\sigma_x ^2 = \sum_{i=1} ^{N} x_i - \bar{x} = \langle \Delta x_i^2
\rangle - \langle \Delta x_i \rangle ^2 $, is calculated through
independent and non-independent trajectory data, in order to
check the accuracy of the calculations of $\sigma_x^2$, both of which
we observe to converge. Therefore, we are able to obtain smooth
$\sigma_x^2$ profiles through averaging over many realisations, which
are of the order of 100 runs per system.

All LJ calculations were performed at the Max Planck Institute for Intelligent Systems
(Stuttgart, Germany). The GF calculations were performed
using the Chebyshev supercomputer of Moscow State University (Moscow,
Russia).

\end{document}